\documentclass[aps,prd,twocolumn,groupedaddress,nofootinbib]{revtex4}
\pdfoutput=1

\usepackage {natbib}
\usepackage {graphicx}
\usepackage {color}

\usepackage{amsmath, amsfonts, amssymb}
\usepackage{hyperref}

\newcommand{\Ord}[1]{\mathcal{O}\left(#1\right)}	

\newcommand{\expect}[1]{E\left[#1\right]}

\newcommand{\dop}{\lambda}
\newcommand{\ddop}{\Delta\dop}
\newcommand{\ddopf}{\Delta\dopf}
\newcommand{\ddopc}{\Delta\dopc}

\newcommand{\dopc}{\co{\dop}}
\newcommand{\dopf}{\ic{\dop}}
\newcommand{\dopsig}{\dop_\sig}

\newcommand{\dopS}{\mathbb{P}}
\newcommand{\dopT}{\mathbb{T}}

\newcommand{\CC}{C}	
\newcommand{\CCc}{\co{\CC}}  
\newcommand{\CCf}{\ic{\CC}}  

\newcommand{\CCtot}{\CC_{\mathrm{tot}}}

\newcommand{\cratio}{\varkappa}

\newcommand{\hrms}{h_{\mathrm{rms}}}
\newcommand{\hrmsk}{h_{\mathrm{rms},k}}

\newcommand{\hth}{h_\thresh}

\newcommand{\prob}[2]{P\left(#1|#2\right)}

\newcommand{\detect}{\mathrm{det}}
\newcommand{\FA}{\mathrm{fA}}
\newcommand{\FD}{\mathrm{fD}}

\newcommand{\pdet}{p_{\detect}}
\newcommand{\pFA}{p_{\FA}}
\newcommand{\pFD}{p_{\FD}}
\newcommand{\erFD}{\beta}	
\newcommand{\erFA}{\alpha}	

\newcommand{\thresh}{\mathrm{th}}

\newcommand{\Nseg}{N}
\newcommand{\tk}{{t_k}}
\newcommand{\Tseg}{\Delta T}	
\newcommand{\Tobs}{T}
\newcommand{\Tmax}{\Tobs_{\mathrm{max}}}

\newcommand{\Nt}{\mathcal{N}} 	
\newcommand{\Ntf}{\ic{\Nt}}	
\newcommand{\Ntc}{\co{\Nt}}	

\newcommand{\F}{\mathcal{F}}
\newcommand{\ideal}{{0}}
\newcommand{\sumF}{\Sigma}
\newcommand{\sumFideal}{\sumF_\ideal}

\newcommand{\Fk}{\F_{k}}

\newcommand{\RHO}{\rho}		
\newcommand{\RHOk}{\RHO_k}
\newcommand{\RHOopt}{\RHO_{\mathrm{opt}}}

\newcommand{\RHOF}{\RHO_{\F}}
\newcommand{\RHOsumF}{\RHO_{\sumF}}
\newcommand{\RHOsumFideal}{\RHO_{\sumFideal}}

\newcommand{\freq}{f}
\newcommand{\fdot}{\dot{\freq}}

\newcommand{\co}[1]{\widetilde{#1}}
\newcommand{\ic}[1]{\widehat{#1}}
\newcommand{\eps}{\varepsilon}

\newcommand{\semi}{\mathrm{sc}}

\newcommand{\FFT}{\mathrm{FFT}}
\newcommand{\SFT}{\mathrm{SFT}}
\newcommand{\sig}{{\mathrm{s}}}
\newcommand{\opt}[1]{{#1}_{\mathrm{opt}}}
\newcommand{\optO}[1]{{#1}^{(0)}_{\mathrm{opt}}}

\newcommand{\refac}{\gamma}
\newcommand{\vol}{\mathcal{V}}

\newcommand{\av}[1]{\bar{#1}}
\newcommand{\avg}[1]{\left\langle #1 \right\rangle}

\newcommand{\Sn}{S_{\mathrm{n}}}

\newcommand{\Lagr}{L}

\newcommand{\nc}{{\co{n}}}
\newcommand{\nf}{{\ic{n}}}

\newcommand{\mis}{\mu}
\newcommand{\misc}{\co{\mis}}
\newcommand{\misf}{\ic{\mis}}
\newcommand{\misfideal}{\misf_\ideal}
\newcommand{\mismax}{m}
\newcommand{\miscmax}{\co{m}}
\newcommand{\misfmax}{\ic{m}}

\newcommand{\gc}{\co{g}}	
\newcommand{\gf}{\ic{g}}	

\newcommand{\etac}{{\co{\eta}}}
\newcommand{\etaf}{{\ic{\eta}}}
\newcommand{\kappac}{{\co{\kappa}}}
\newcommand{\kappaf}{{\ic{\kappa}}}
\newcommand{\deltac}{{\co{\delta}}}
\newcommand{\deltaf}{{\ic{\delta}}}
\newcommand{\epsc}{{\co{\eps}}}
\newcommand{\epsf}{{\ic{\eps}}}

\newcommand{\dev}{w}	
\newcommand{\WSG}{{\mathrm{WSG}}}

\newcommand{\crit}{a}
\newcommand{\critc}{\co{\crit}}
\newcommand{\critf}{\ic{\crit}}

\newcommand{\detCoefs}{D}

\newcommand{\thick}{\theta}

\newcommand{\erf}{\mathrm{erf}}
\newcommand{\erfc}{\mathrm{erfc}}
\newcommand{\erfcInv}{\erfc^{-1}}

\newcommand{\pacfactor}{\xi}

\newcommand{\stageo}{{(0)}}
\newcommand{\stageI}{{(1)}}
\newcommand{\stageII}{{(2)}}
\newcommand{\stagei}{{(i)}}

\newcommand{\stageip}{{(i+1)}}
\newcommand{\stagen}{{(m)}}

\newcommand{\cand}{{\mathrm{cand}}}
\newcommand{\Ncand}{\Nt_\cand}

\newcommand{\dl}{\Delta\deltac}

\newcommand{\days}{\mathrm{days}}
\newcommand{\Hz}{\mathrm{Hz}}
\newcommand{\years}{\mathrm{y}}
\newcommand{\hours}{\mathrm{h}}
\newcommand{\secs}{\mathrm{s}}

\newcommand{\rtHz}{\sqrt{\Hz}}

\newcommand{\Ndet}{N_\mathrm{det}}

\def\commitID{commitID: 5b2a8b2c4f2dd79b729dbde0b35e307ba032a5cc}
\def\commitDATE{ Thu May 3 17:20:34 2012 +0200}

\def\commitSTATUS{CLEAN}

\newcommand{\dcc}{LIGO-P1100156-v5}

\newcommand{\ltitle}{Search for Continuous Gravitational Waves: Optimal StackSlide method at fixed computing cost}
\newcommand{\stitle}{Optimal StackSlide method at fixed computing cost}

\hypersetup{
  pdftitle={\ltitle},
  pdfauthor={R. Prix, M. Shaltev},
  pdfkeywords={\$\commitID-\commitSTATUS{} \$}
}

\begin{document}

\title[\stitle]{\ltitle}
\author{R.\ Prix, M.\ Shaltev}
\affiliation{Albert-Einstein-Institut, Callinstr.\ 38, 30167 Hannover, Germany}
\date{\commitDATE\\\mbox{\dcc}}

\begin{abstract}
Coherent wide parameter-space searches for continuous gravitational waves are typically
limited in sensitivity by their prohibitive computing cost. Therefore semi-coherent methods
(such as StackSlide) can often achieve a better sensitivity.
We develop an analytical method for finding \emph{optimal} StackSlide parameters at
fixed computing cost under ideal conditions of gapless data with Gaussian stationary noise.
This solution separates two regimes: an \emph{unbounded} regime, where it is always optimal to use
all the data, and a \emph{bounded} regime with a finite optimal observation time.
Our analysis of the sensitivity scaling reveals that both the fine- and coarse-grid mismatches
contribute equally to the average StackSlide mismatch, an effect that had been overlooked in
previous studies.
We discuss various practical examples for the application of this optimization framework,
illustrating the potential gains in sensitivity compared to previous searches.
\end{abstract}

\pacs{XXX}

\maketitle

\section{Introduction}
\label{intro}


\textit{Motivation.}
The detection of continuous gravitational waves (CWs) from spinning neutron stars (NSs) in our galaxy
remains an elusive goal, despite the global network of detectors LIGO \cite{LIGORef:2009}, Virgo
\cite{VirgoRef:2011} and GEO\,600 \cite{GEORef:2010} having completed their
initial and enhanced science runs (e.g.\ see \cite{powerfluxS5,2010ApJ...713..671A,Abbott:2009nc,2010ApJ...722.1504A}).
The search for CWs will likely remain a difficult challenge with uncertain prospects even in the era
of Advanced detectors \cite{AdvLIGORef:2010,AdvVirgoRef:2009,GeoHFRef:2006} and
third-generation detectors such as ET \cite{ETRef:2010}.
Two main reasons for this are (i) astrophysical priors on CWs and
(ii) the large parameter space of signal parameters to explore (cf.\ \cite{prix06:_cw_review} for a
review and further references).

(i) Current astrophysical priors contain large uncertainties on the expected strength of CW
emissions from spinning NSs, with a strong bias towards extremely weak signals, informed by
the population of known pulsars as well as by a statistical analysis of a putative galactic ``gravitar''
population \cite{2008PhRvD..78d4031K}.
(ii) The required number of templates for a coherent matched-filter search over a range of unknown signal
parameters typically grows very rapidly with increasing  duration of data analyzed. Therefore only a
fraction of the available data can be analyzed coherently
(e.g.\ see \cite{Brady:1997ji,2008CQGra..25w5011W,lsc06:_coher_scorp_x}).

It was realized early on \cite{Brady:1998nj} that in situations where the total computing cost of
the search is constrained, a \emph{semi-coherent} approach could typically achieve better
sensitivity than coherent matched filtering:
shorter segments of data are analyzed coherently, then the statistics from these segments are
combined incoherently. One method of incoherent combination simply consists of summing the
statistics from the different segments, which is typically referred to as the ``StackSlide''
method in the CW context (also known as the Radon transform).
The template bank used for the semi-coherent combination of coherent statistics is referred to as the
\emph{fine grid}, as it typically requires a higher resolution than the template banks of the
per-segment coherent searches (referred to as \emph{coarse grids}).
Details of the respective template banks will be discussed in Sec.~\ref{sec:template-counting}.

There are a number of different semi-coherent methods: for example, recent work
\cite{pletsch2011:_sliding} has shown that StackSlide sensitivity can be improved by a
sliding coherence-window approach.
A closely related variant to StackSlide is the \emph{Hough transform} \cite{Krishnan:2004sv}, which counts
the number of segments in which the statistic crosses a given threshold, instead of summing the statistics.
This is generally less
sensitive, but is designed to be more robust in the presence of strong non-stationarities.
A somewhat different semi-coherent approach are cross-correlation methods, described in more detail
in \cite{2008PhRvD..77h2001D}.

Related to the semi-coherent methods are the so-called \emph{hierarchical} schemes, which
consist of following up ``promising'' candidates from a (coherent or semi-coherent) search
by subsequent, more sensitive searches, referred to as ``stages''. This procedure is iterated until
the parameter space of surviving candidates is
sufficiently narrowed down for a fully coherent follow-up using
all the data. Work on implementing such schemes in practice is still ongoing.

\textit{Optimization problem.}
In this paper we focus on the standard single-stage StackSlide method, which was also
used in previous optimization studies \cite{Brady:1998nj,PhysRevD.72.042004}, and is relatively
straightforward to describe analytically.

Any search method contains a number of tuneable parameters, such as the template-bank mismatch, the
data selection procedure, and the number and length of segments to analyze.
Hierarchical schemes would further require specification of the number of stages and the respective
distributions of computing costs and candidate thresholds.
The sensitivity of a search generally depends on all these choices, and we therefore need to
study how to maximize sensitivity as a function of these parameters.

This optimization problem has been studied previously by
Brady and Creighton~\cite{Brady:1998nj} (henceforth 'BC') and subsequently by
Cutler, Gholami and Krishnan~\cite{PhysRevD.72.042004} (in the following 'CGK').
Both studies have focused on the wider problem of optimizing a multi-stage hierarchical scheme
of StackSlide stages, and have directly resorted to fully numerical exploration of the optimization
problem. Here instead we focus on the simpler single-stage search, which allows us to fully
analytically analyze the problem. In the next step this can be used as a building block to attack
the optimization of hierarchical schemes.

Note that for a network of detectors with different noise-floors, the choice of detectors to use at
fixed computing cost is part of the optimization problem, but under the present assumption
of ``ideal data'' the answer can be obtained independently \cite{2007gr.qc.....2068P}.
More work is required to develop a practical algorithm to compute the optimal search
parameters for given data from a network of detectors, including gaps, non-stationarities and
various detector artifacts.

\textit{Summary of main results.}
Careful analysis of the sensitivity scaling shows that the average
StackSlide mismatch is given by the \emph{sum} of the average mismatches from the coarse- and fine-grid
template banks, an effect that had previously been overlooked.
Note that we allow for independent coarse- and fine-grid mismatches, while BC and CGK forced them to
be equal as an \emph{ad-hoc} constraint.

The analytic optimization is achieved by using local power-law approximations to the computing-cost
and sensitivity functions. The results provide analytic self-consistency conditions for the optimal
solution: if the initial power-law coefficients agree with those found at the
analytic solution, then the solution is self-consistent and (locally) optimal.
If this is not the case, one can iterate over successive solutions or scan a range of
StackSlide parameters, in order to ``bootstrap'' into a self-consistent optimal solution.

We find that the analytic solution for StackSlide searches separates two
different regimes depending on the power-law coefficients: a \emph{bounded} regime in which there is
a finite optimal observation time, and an \emph{unbounded} regime in which the optimal solution
always consists of using all of the available data, irrespective of the available computing-cost.
\textit{Plan of the paper.}
In Sec.~\ref{sec:maxim-chance-detect} we discuss the general CW optimization problem, which includes
the single-stage StackSlide search as the lowest-level building block.
In Sec.~\ref{sec:prop-single-stage} we derive the sensitivity estimate and computing-cost functions
for StackSlide searches, and motivate their approximation as local power-laws.
After deriving in Sec.~\ref{sec:optim-sens-stacksl} the general analytical solution and discussing a
few special cases, we provide examples for the practical application of this
framework in Sec.~\ref{sec:exampl-appl}: directed searches, all-sky searches, and searches
for CWs from NSs in binary systems.

\section{Maximizing probability of a CW detection}
\label{sec:maxim-chance-detect}

The goal for wide parameter-space CW searches for unknown signals should be to maximize the
probability of detection, given current astrophysical priors, detector data, and finite computing resources.
Conceptually one can think of this problem as a hierarchy of two questions:
\begin{enumerate}
\item[(i)] What parameter-space $\dopS\subseteq\dopS^{(0)}$ to search? More generally: how to distribute
  the total available computing power $\CC_0$ over the space $\dopS^{(0)}$ of possible CW signals,
  given astrophysical priors, detector data and an (optimal) search method?

\item[(ii)] What is the optimal search method? Namely, which method yields the highest detection
  probability $\pdet$ on a parameter-space cell $\Delta\dopS$ if we spend computing-cost $\Delta\CC$ on it?
\end{enumerate}
The answer to the first question relies on the second, but we can analyze the lower-level second
question independently of the first. There has been surprisingly little work on this problem so far.
The first question has hardly been addressed at all, except for recent work by Knispel
\cite{knispel2011:_thesis}. The second question has been studied previously by
BC~\cite{Brady:1998nj} and CGK~\cite{PhysRevD.72.042004}, assuming a specific type of hierarchical
scheme, which we refer to as the \emph{classical hierarchical scheme} (CHS).

In the CHS one performs a hierarchy of semi-coherent searches (called \emph{stages}), starting
with a relatively low-sensitivity search over the whole initial parameter space $\dopS^\stageI$.
Promising candidates crossing the first-stage threshold are selected and constitute the search
subspace $\dopS^\stageII\subset\dopS^\stageI$ for the second, higher-sensitivity stage. This is
iterated until eventually after $m$ such stages a fully-coherent search over all the data can be
performed on the surviving candidates. At this point one has reached the maximal possible
sensitivity for a small portion $\dopS^\stagen\subset\dopS^\stageI$ of the initial parameter space.

Each stage $\stagei$ is characterized by its input parameter-space $\dopS^\stagei$,
a computing-cost constraint $\CC_0^\stagei$ and a false-alarm probability $\pFA^\stagei$.
Each stage selects a candidate subspace $\dopS^\stageip \subset \dopS^\stagei$ to follow up in the next stage.
An optimal per-stage search would result in the highest detection probability
$\pdet^\stagei$ for given signal strength $\hrms$ and constraints $\{\pFA^\stagei,\,\dopS^\stagei,\,\CC_0^\stagei\}$.
The tuneable CHS parameters are therefore the number $m$ of stages and the per-stage constraints
$\{\pFA^\stagei, \CC_0^\stagei\}$. These can be varied in order to maximize the overall detection
probability $\pdet(\hrms)$ for the given total signal parameter-space $\dopS^\stageo$, computing cost $\CC_0 = \sum_{i=1}^m\CC_0^\stagei$
and false-alarm probability $\pFA = \prod_{i=1}^{m}\,\pFA^\stagei$.

This formulation of the CHS suggests that each stage $\stagei$ could be considered an independent
optimization problem for given external constraints $\{\pFA^\stagei,\dopS^\stagei,\CC_0^\stagei\}$,
if none of its internal parameters interfere with the overall hierarchical scheme.
One might contend that the parameter-space resolution of the search violates this clean
factorization: the follow-up space $\dopS^\stageip$ from stage
$\stagei$ depends on its parameter-space resolution, which might impact the required computing-cost
of the next stage.
However, it is easy to see that (to first order) such a coupling is not expected.
The number of candidates $\Ncand$ returned from any stage (except for the last one) will be
dominated by the number $\Nt_\FA$ of false alarms. Therefore
$\Ncand \approx \Nt_\FA \approx \pFA\,\Nt$, where $\Nt$ is the number of (approximately) independent
templates searched in this stage. This can be estimated as $\Nt\approx V_\dopS /v_0$,
in terms of the (metric) volume $V_\dopS$ of the parameter space $\dopS$, and the volume $v_0$
covered by one template.
Therefore the \emph{number} $\Ncand$ of follow-up candidates from any stage does indeed depend on
its parameter-space resolution, which depends on the internal stage parameters.
However, the computing cost of the next stage depends
primarily on the \emph{volume} of the follow-up parameter-space, which is
$V_\FA \approx \Nt_\FA\,v_0 \approx \pFA\,V_\dopS$,
and is therefore \emph{independent} of internal stage parameters.
It is interesting to note that each stage $\stagei$ in this scheme achieves a reduction of the input
parameter-space volume by roughly a factor of the false-alarm probability $\pFA^\stagei$, irrespective of
the internal details of that search.

The optimal per-stage search method is essentially unknown, but following BC and CGK we focus on a
known good strategy, namely the StackSlide method.
While different semi-coherent methods differ in the details and their exact sensitivity, they share
the main characteristics of coherently searching $\Nseg$ shorter segments of length $\Tseg$, and
combining them incoherently in some way.
We roughly expect the \emph{sensitivity per cost} of different methods to behave qualitatively
similarly to the StackSlide method, but more work would be required to study this in detail.

\section{Properties of a single-stage StackSlide search}
\label{sec:prop-single-stage}

The general StackSlide scheme consists of dividing the data (of total duration $\Tobs$) into $\Nseg$
segments of duration $\Tseg = \Tobs/\Nseg$, then performing a coherent matched-filter search on each
segment and combining these statistics \emph{incoherently} to a new statistic $\sumF$ by summing them
across segments. The coherent matched-filter statistic used is the $\F$-statistic, which was first derived in
\cite{jks98:_data} and extended to multiple detectors in \cite{cutler05:_gen_fstat}.
Using the same amount of data as a fully coherent search, the resulting semi-coherent statistic is
less sensitive, but substantially cheaper to compute over a wide parameter space.
At \emph{fixed computing cost} a semi-coherent search is therefore generally more sensitive than
a fully coherent $\F$-statistic search.

\emph{Notation:} we distinguish quantities $Q$ that can refer to either the
coherent or the incoherent step in the following way: we use a tilde, i.e.\ $\co{Q}$ when referring
to the coherent step, and a hat, i.e.\ $\ic{Q}$ when referring to the incoherent step.
For the following derivations we restrict ourselves to a single-detector formalism for simplicity,
but we state the (trivial) generalization to $\Ndet\ge1$ detectors of relevant results.

\subsection{The StackSlide search method}
\label{sec:stack-slide-search}

Let $k=1\ldots\Nseg$ be the index over segments, and $\dop \in \dopS$ a point in
the search space $\dopS$ of signal parameters.
The ``ideal'' StackSlide statistic $\sumFideal(\dop)$ is defined as
\begin{equation}
  \label{eq:7}
  \sumFideal(\dop) \equiv \sum_{k=1}^N 2\Fk(\dop)\,,
\end{equation}
i.e.\ a simple sum of $\F$-statistic values $\{2\Fk(\dop)\}_{k=1}^\Nseg$ computed at \emph{the same}
template point $\dop$ across all $\Nseg$ segments.

This would require computing the $\F$-statistic over the same template bank as $\sumFideal$ in every
segment. However, the metric resolution of $\sumFideal$ is generally finer than that of the
single-segment $\F$-statistics \cite{Pletsch:2010a}, and therefore this approach would spend
unnecessary computing cost on the coherent $\F$-statistic.
In practice $\F$ is therefore computed over a \emph{coarse grid} of $\co{\Nt}$
templates in each segment $k$, and is interpolated in order to sum $\F$ on the
\emph{fine grid} of $\Ntf \ge \Ntc$ templates (e.g.\ see \cite{2009PhRvL.103r1102P}).

Typically the interpolation consists of picking the \emph{closest} (in the metric sense) coarse-grid
point $\dopc_k(\dopf)$ to the fine-grid point $\dopf$ from every segment $k$, i.e.\ we approximate Eq.~(\ref{eq:7}) by
\begin{equation}
  \label{eq:8}
  \sumF(\dopf) \equiv \sum_{k=1}^N 2\Fk\left(\dopc_k(\dopf)\right) \approx \sumFideal(\dopf)\,,
\end{equation}
which we refer to as the ``interpolating'' StackSlide statistic $\sumF$. The following sensitivity
optimization focuses exclusively on this interpolating StackSlide method, which is the
most relevant approach for current practical applications.
The subtle difference between \emph{interpolating StackSlide} $\sumF$ and \emph{ideal StackSlide} $\sumFideal$
with respect to its sensitivity and mismatches has been overlooked in previous studies, and will be
important for the optimization problem.

\subsection{Mismatch and metric}
\label{sec:mismatch-metric}

\subsubsection{$\F$-statistic mismatch}
\label{sec:coherent-per-segment}

In the presence of a signal timeseries $s(t,\dopsig)$ with phase parameters $\dopsig$, the
statistic $2\Fk(\dopc)$ in a point $\dopc$ follows a non-central $\chi^2$-distribution with
four degrees of freedom and non-centrality parameter $\RHOk^2(\dopsig,\dopc)$. We denote this
probability distribution as
\begin{equation}
  \label{eq:20}
  \prob{2\F_k}{\RHOk^2} = \chi^2_4(2\F_k;\RHOk^2)\,,
\end{equation}
which has the expectation value
\begin{equation}
  \label{eq:10}
  \expect{2\Fk(\dopsig,\dopc)} = 4 + \RHOk^2(\dopsig,\dopc)\,.
\end{equation}
The quantity $\RHOk$ is often referred to as the coherent signal-to-noise ratio (SNR).
In the case of a perfectly-matched template $\dopc = \dopsig$, the resulting ``optimal''
SNR \cite{jks98:_data} in segment $k$ can be expressed as
\begin{align}
    \RHOk^2(\dopsig,\dopsig) &= \frac{2}{\Sn} \int_\tk^{\tk+\Tseg} s^2(t,\dopsig)\,d t \notag\\
    &\equiv \frac{2}{\Sn}\,\hrmsk^2\,\Tseg\,,   \label{eq:84}
\end{align}
where $t_k$ is the start-time of the $k$th segment, $\Sn$ is the (single-sided) noise power spectral
density at the signal frequency $f_\sig$. In the second equality we defined the rms signal strength
$\hrmsk$ \cite{PhysRevD.72.042004} in segment $k$, which is a useful measure of the \emph{intrinsic}
signal strength in the detectors, independently of the quality and the amount of data used.

The signal strength $\hrmsk$ depends on the intrinsic signal amplitude $h_0$, the sky-position, polarization
angles, and detector orientation during segment $k$. One can show \cite{jks98:_data,prix:_cfsv2}
that averaging $\hrmsk^2$ isotropically over sky-positions and polarization angles yields the
relation
$\langle\hrmsk^2\rangle_{\mathrm{sky,pol}}=(2/25)\,h_0^2$.
Furthermore, for segment lengths of order $\Tseg\gtrsim \Ord{1\,\days}$, the averaging in
Eq.~\eqref{eq:84} results in $\hrmsk$ tending towards a constant over all segments. Therefore it
will be convenient to approximate $\hrmsk \approx \hrms$, and so we can write
\begin{equation}
  \label{eq:90}
  \RHOk^2(\dopsig,\dopsig) \approx \frac{2}{\Sn}\hrms^2\,\Tseg \equiv \RHOopt^2(\Tseg)\,,
\end{equation}
which defines the average optimal SNR $\RHOopt$ for given segment length $\Tseg$.

Note that this approximation only applies to the perfectly-matched SNR $\RHOk(\dopsig,\dopsig)$.
The ``mismatched'' SNR $\RHOk(\dopsig,\dopc)$ in an offset template $\dopc = \dopsig + \ddop$ is
reduced with respect to the optimal SNR $\RHOopt(\Tseg)$.
The corresponding relative loss defines the (segment-specific) \emph{mismatch} function
$\misc_k(\dopsig,\dopc)$, namely
\begin{align}
  \misc_k(\dopsig,\dopc) &\equiv 1 - \frac{\RHOk^2(\dopsig,\dopc)}{\RHOopt^2(\Tseg)} \notag\\
  &= \gc_{ij,k}(\dopsig)\,\ddop^i\,\ddop^j + \Ord{\ddop^3}\,,  \label{eq:93}
\end{align}
where Taylor-expansion for small offsets $\ddop$ defines the (coherent) metric
tensor $\gc_{ij,k}(\dop)$ for segment $k$.
The concept of the parameter-space metric was first introduced in
\cite{bala96:_gravit_binaries_metric,owen96:_search_templates}, and analyzed in the context of
a simplified CW statistic \cite{Brady:1997ji} and the $\F$-statistic \cite{prix06:_searc}.

The per-segment \emph{coarse-grid} template bank is constructed under the constraint that no signal
point $\dopsig\in\dopS$ should exceed a given \emph{maximal mismatch} $\miscmax$ to its closest
(i.e.\ with the smallest mismatch) coarse-grid template $\dopc_k(\dopsig)$, namely
\begin{equation}
  \label{eq:114}
  \misc_k\left(\dopsig,\dopc_k(\dopsig)\right) \le \miscmax\,\quad\mbox{for all}\quad \dopsig\in\dopS\,.
\end{equation}

\subsubsection{Mismatch of ``ideal'' StackSlide}
\label{sec:ideal-stackslide}

The ``ideal'' StackSlide statistic $\sumFideal$ defined in Eq.~(\ref{eq:7}) is the basis for the
definition of the semi-coherent metric
\cite{Brady:1998nj,Pletsch:2010a,2011arXiv1109.0501M}).
The statistic $\sumFideal$ follows a non-central $\chi^2$-distribution with $4\Nseg$ degrees of
freedom, denoted as
\begin{equation}
  \label{eq:44}
  \prob{\sumFideal}{\RHOsumFideal^2} = \chi^2_{4\Nseg}(\sumFideal;\RHOsumFideal^2)\,,
\end{equation}
with non-centrality parameter
\begin{equation}
  \label{eq:72}
  \RHOsumFideal^2(\dopsig,\dopf) \equiv \sum_{k=1}^\Nseg \RHOk^2(\dopsig,\dopf)\,,
\end{equation}
where $\dopsig$ are the signal parameters and $\dopf$ is the location of a fine-grid template. The
corresponding expectation value is
\begin{equation}
  \label{eq:11}
  \expect{\sumFideal(\dopsig,\dopf)} = 4 \Nseg + \RHOsumFideal^2\,.
\end{equation}
The perfectly-matched non-centrality parameter $\RHOsumFideal^2(\dopsig,\dopsig)$ can be expressed as
\begin{align}
    \RHOsumFideal^2(\dopsig,\dopsig) &= \sum_{k=1}^\Nseg \RHOk^2(\dopsig,\dopsig) \notag\\
    & = \Nseg\RHOopt^2(\Tseg) \notag\\
    & = \RHOopt^2(\Tobs)\,,   \label{eq:28}
\end{align}
which is identical to that of a perfectly-matched $\F$-statistic over the same total duration
$\Tobs$, as seen from Eq.~(\ref{eq:90}). The reason why the StackSlide statistic $\sumFideal$ is less
sensitive than the $\F$-statistic for the same amount of data $\Tobs$ stems from the different
degrees of freedom of the respective distributions, namely $\chi^2_4(\RHO^2)$ for the
$\F$-statistic as opposed to $\chi^2_{4\Nseg}(\RHO^2)$ for StackSlide $\sumFideal$.

The mismatch function $\misfideal(\dopsig,\dopf)$ of ideal StackSlide is defined in analogy to Eq.~(\ref{eq:93}) as
\begin{align}
  \misfideal(\dopsig,\dopf) &\equiv 1 - \frac{\RHOsumFideal^2(\dopsig,\dopf)}{\RHOopt^2(\Tobs)} \notag\\
  & = \gf_{ij}(\dopsig)\,\ddop^i\,\ddop^j + \Ord{\ddop^3}\,,  \label{eq:92}
\end{align}
where $\ddop \equiv \dopf - \dopsig$ is the offset between the fine-grid template $\dopf$ and the signal location
$\dopsig$, and Taylor-expansion in small $\ddop$ defines the (semi-coherent) metric tensor $\gf_{ij}$.
Using Eqs.~(\ref{eq:28}) and (\ref{eq:93}), we can rearrange the expression for the mismatch as
\begin{align}
  \misfideal(\dopsig,\dopf) &= \frac{1}{\Nseg}\sum_{k=1}^\Nseg \misc_k(\dopsig,\dopf)  \notag\\
  &\approx \left( \frac{1}{\Nseg} \sum_{k=1}^\Nseg \gc_{ij,k}(\dopsig)\right)\,\ddop^i \ddop^j\,,  \label{eq:96}
\end{align}
which shows that the ideal StackSlide mismatch $\misfideal$ and metric $\gf_{ij}$ are segment-averages of
the coherent mismatches and metrics, respectively.

The \emph{fine-grid} template bank of a StackSlide search is constructed under the constraint that
no signal point $\dopsig\in\dopS$ should exceed a given \emph{maximal mismatch} $\misfmax$ to its
\emph{closest} (i.e.\ with the smallest mismatch) fine-grid template $\dopf(\dopsig)$, namely
\begin{equation}
  \label{eq:115}
  \misfideal\left(\dopsig,\,\dopf(\dopsig)\right) \le \misfmax\,\quad\mbox{for all}\quad \dopsig\in\dopS\,.
\end{equation}

\subsubsection{Mismatch of ``interpolating'' StackSlide}
\label{sec:real-stackslide}

We can now combine the above results to derive the mismatch of the interpolating StackSlide
statistic $\sumF$ defined in Eq.~(\ref{eq:8}). This statistic follows a non-central
$\chi^2_{4\Nseg}$ distribution, namely
\begin{equation}
  \label{eq:94}
  \prob{\sumF}{\RHOsumF^2} = \chi^2_{4\Nseg}(\sumF;\RHOsumF^2 )\,,
\end{equation}
with non-centrality parameter
\begin{equation}
  \label{eq:100}
  \RHOsumF^2(\dopsig,\dopf) \equiv \sum_{k=1}^\Nseg \RHOk^2\left(\dopsig,\dopc_k(\dopf) \right)\,,
\end{equation}
where $\dopsig$ are the signal phase parameters, and $\dopc_k(\dopf)$ is the closest coarse-grid template
in segment $k$ to the fine-grid point $\dopf$.

\begin{figure}[htbp]
  \centering
  \includegraphics[width=\columnwidth]{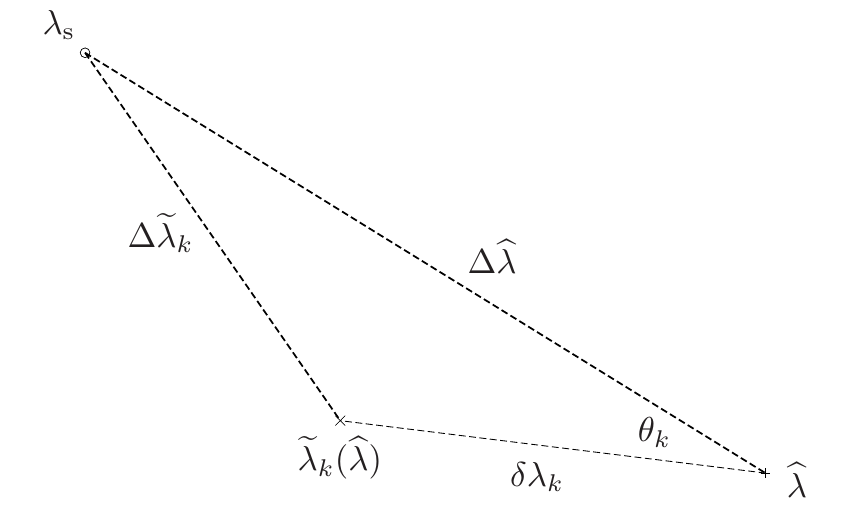}
  \caption{Mismatch triangle formed by the signal point $\dopsig$, closest
    fine-grid template $\dopf$, and the coarse-grid template $\dopc_k(\dopf)$ closest to $\dopf$ in segment $k$.}
  \label{fig:mismatchTriangle}
\end{figure}

The mismatch function $\misf(\dopsig,\dopf)$ of interpolating StackSlide is therefore
\begin{align}
  \misf(\dopsig,\dopf) &\equiv 1- \frac{\RHOsumF^2(\dopsig,\dopf)}{\RHOopt^2(\Tobs)} \notag\\
  &= \frac{1}{\Nseg}\sum_{k=1}^{\Nseg} \misc_k\left( \dopsig,\,\dopc_k(\dopf)\right)\,,   \label{eq:108}
\end{align}
which allows us to express the mismatched non-centrality parameter as
\begin{equation}
  \label{eq:112}
  \RHOsumF^2(\dopsig,\dopf) = \left(1 - \misf(\dopsig,\dopf)\right)\,\RHOopt^2(\Tobs)\,.
\end{equation}
The extra offset per-segment, $\delta\dop_k \equiv \dopc_k(\dopf) - \dopf$, incurred due to using the
closest coarse-grid point $\dopc_k(\dopf)$ instead of the fine-grid point $\dopf$ tends to increase
the mismatch with respect to the ideal mismatch function $\misfideal$ of Eq.~(\ref{eq:96}).
In order to quantify this effect, we write the effective per-segment offset
from a signal as $\ddopc_k \equiv \dopc_k(\dopf) - \dopsig$, while the ideal per-segment offset
would be $\ddopf \equiv \dopf - \dopsig$.
We can write $\ddopc_k = \ddopf + \delta\dop_k$, and inserting this into the coherent-metric
of Eq.~(\ref{eq:93}) we obtain (neglecting higher-order terms $\Ord{\Delta\dop^3}$):
\begin{multline}
  \label{eq:109}
    \misc_k\left(\dopsig,\dopc_k(\dopf)\right) = \gc_{ij,k}\,\ddopc_k^i\ddopc_k^j\\
    = \misc_k(\dopsig,\dopf) + \gc_{ij,k}\, \delta\dop_k^i \,\delta\dop_k^j
    + 2\gc_{ij,k}\,\ddopf^i \,\delta\dop_k^j\,,
\end{multline}
where in the first term we recover the ideal per-segment mismatch function of Eq.~(\ref{eq:96}), the
second term represents an extra mismatch due to the offset $\delta\dop_k$, while the last term
depends on the opening angle $\theta_k$ of the mismatch triangle (see Fig.~\ref{fig:mismatchTriangle}),
namely $2|\ddopf||\delta\dop_k|\cos\theta_k$, with
mismatch norm defined as $|x|^2\equiv \gc_{ij,k}\,x^i x^j$.

We assume that the fine-grid point $\dopf$ falls randomly into the Wigner-Seitz cell of the closest
coarse-grid template $\dopc_k(\dopf)$ in segment $k$. Given that the coarse-grid metric $\gc_{ij,k}$ generally varies
across segments, we further assume that the offset $\delta\dop_k$ approximates a uniform random
sampling of the coarse-grid Wigner-Seitz cell.
Inserting Eq.~(\ref{eq:109}) into (\ref{eq:108}), we see that the average over the angle-term
$|\ddopf||\delta\dop_k|\cos\theta_k$ will tend to zero, as any sign of $\cos\theta_k$ is equally likely,
while the average norm $|\delta\dop_k|^2$ will tend to the average mismatch $\avg{\misc}$ of the
coarse-grid template bank, and so we obtain
\begin{equation}
  \label{eq:110}
  \misf(\dopsig,\dopf) \approx \misfideal(\dopsig,\dopf) + \avg{\misc}\,.
\end{equation}
When estimating the sensitivity of the interpolating StackSlide statistic, we will further average this expression
over randomly-chosen signal locations $\dopsig$, and therefore the above approximate averaging
expressions will become exact, and we obtain
\begin{equation}
  \label{eq:111}
  \avg{\misf} = \avg{\misfideal} + \avg{\misc}\,,
\end{equation}
where averaging is performed over the coarse- and fine-grid template banks (i.e.\ the
respective Wigner-Seitz cells).

The probability distribution of signal mismatches in a given template bank constructed
with a certain maximal mismatch $\mismax$ depends on the
structure and dimensionality of the template bank. The corresponding average mismatch can be
expressed as $\avg{\mis} = \pacfactor\, \mismax$, where $\pacfactor \in (0,1)$ is a characteristic geometric factor of the template bank.
Such mismatch distributions were studied quantitatively, for example in \cite{2009PhRvD..79j4017M}.
For hyper-cubic lattices, the geometric relation is well known to be exactly $\avg{\mis} = \mismax/3$,
which was used in previous optimization studies \cite{Brady:1998nj,PhysRevD.72.042004}.
For the more efficient $A_n^*$-lattices this geometric factor is approximately $\pacfactor\approx 0.5 - 0.6$
for low dimensions $n\lesssim 6$.
Here we allow for general geometric factors $\pacfactor$, but for simplicity we
assume it to be identical for the fine- and coarse-grid template banks, and so Eq.~(\ref{eq:111})
can be written as
\begin{equation}
  \label{eq:39}
  \avg{\misf} = \pacfactor ( \misfmax + \miscmax)\,,\quad \mbox{with}\quad \pacfactor \in (0,1)\,,
\end{equation}
where $\misfmax$ and $\miscmax$ are the maximal mismatch parameters of fine- and coarse-grid
template banks, respectively.

Averaging the non-centrality parameter $\RHOsumF^2$ of Eq.~(\ref{eq:112})
over random signal parameters $\dopsig$ at fixed signal strength $\hrms$, we can now obtain the
expression
\begin{equation}
  \label{eq:113}
  \avg{\RHOsumF^2} = \left[ 1 - \pacfactor(\misfmax + \miscmax)\right]\,\frac{2\Ndet}{\Sn}\,\hrms^2\,\Tobs\,,
\end{equation}
where we (trivially) generalized the result to the case of a network of $\Ndet$ detectors.
In this case $\Sn$ refers to the harmonic mean over individual-detector PSDs, and $\hrms$ is a
noise-weighted average over rms-amplitudes from different detectors (e.g.\ see \cite{prix:_cfsv2}).
The fact that \emph{both} the coarse- and fine-grid mismatches enter this expression
has been overlooked in previous studies \cite{Brady:1998nj,PhysRevD.72.042004}, where only the
fine-grid mismatch $\misfmax$ had been included\footnote{These studies additionally imposed the
  \emph{ad-hoc} constraint of \mbox{$\miscmax = \misfmax$} in the computing-cost expressions}.

\subsection{Sensitivity estimate}
\label{sec:sens-estim-mism}

The false-alarm and false-dismissal probabilities for a given threshold $\sumF_\thresh$ of the
StackSlide statistic $\sumF$ of Eq.~(\ref{eq:8}) are
\begin{align}
  \pFA(\sumF_\thresh)          &= \int_{\sumF_\thresh}^\infty \chi^2_{4\Nseg}(\sumF;0)\,d \sumF\,,   \label{eq:97a}\\
  \pFD(\sumF_\thresh,\RHOsumF^2) &= \int_{-\infty}^{\sumF_\thresh} \chi^2_{4\Nseg}(\sumF;\RHOsumF^2)\,d \sumF\,,  \label{eq:97b}
\end{align}
where the special case of a coherent $\F$-statistic search corresponds to $\Nseg=1$.

Sensitivity is often quantified in terms of the weakest (rms-) signal strength $\hth$ required to obtain a given
detection probability $\pdet^* = 1 - \pFD^*$ at a given false-alarm probability $\pFA^*$.
This requires inverting Eq.~(\ref{eq:97a}) to obtain the critical threshold $\sumF_\thresh^* = \sumF_\thresh(\pFA^*)$,
then substituting this into Eq.~(\ref{eq:97b}) and inverting to find the critical non-centrality parameter
\begin{equation}
  \label{eq:81}
  \RHOsumF^{*2} = \RHOsumF^2(\pFA^*,\pFD^*,\Nseg)\,.
\end{equation}

The signal location $\dopsig$ is generally unknown, therefore the mismatch $\misf(\dop_\sig,\dopf)$
of the closest template $\dopf$ and the corresponding mismatched non-centrality parameter
$\RHOsumF^2(\dop_\sig,\dopf)$ of Eq.~(\ref{eq:112}) follow a random distribution.
In order to estimate the threshold rms signal strength $\hth$, one would have to compute
$\pFD(\pFA^*,\hth)$ by \emph{averaging} the right-hand side of
Eq.~(\ref{eq:97b}) over the (known) mismatch distribution of $\misf$.
Furthermore, for statements about physical upper limits and sensitivity of a given search pipeline, it is often
required to quantify the sensitivity in terms of the \emph{intrinsic} GW amplitude $h_0$, instead
of the rms detector strain $\hrms$, which would require further averaging of Eq.~(\ref{eq:97b}) over
the (potentially) unknown sky-position and polarization parameters. This problem has recently been
studied in detail in \cite{wette2011:_sens}.

For our present purpose it will be sufficient to obtain the correct \emph{scaling} of
sensitivity with StackSlide parameters $\{\Nseg,\Tobs,\miscmax,\misfmax\}$, while the absolute sensitivity level is less important.
We will therefore employ the usual simplification of this problem, which consists in averaging $\RHOsumF^2$
instead of $\pFD(\RHOsumF^2)$ over the mismatch distribution of $\misf$, so we approximate
\begin{align}
  \pFD(\pFA^*,\hth) &= \avg{\left.\pFD(\pFA^*,\RHOsumF^2)\right|_{\hth}}_{\dopsig} \notag\\
  & \approx \pFD\left(\pFA^*,\avg{\RHOsumF^2}_{\dopsig}\right)\,.  \label{eq:117}
\end{align}
The results of \cite{wette2011:_sens} indicate that this indeed approximately preserves the
\emph{scaling} of sensitivity as a function of StackSlide parameters.

We can now use Eq.~(\ref{eq:113}) to translate the critical non-centrality parameter
$\RHOsumF^{*2}$ of Eq.~(\ref{eq:81}) into a threshold rms signal-strength $\hth$, namely
\begin{equation}
  \label{eq:99}
  \hth^{-2} = \frac{2\Ndet}{\RHOsumF^{*2}} \,\left[ 1 - \pacfactor(\misfmax + \miscmax)\right]\,\frac{\Tobs}{\Sn}\,.
\end{equation}
Following the Neyman-Pearson criterion we want to maximize detection probability
$\pdet^*=1-\pFD^*$ at fixed false-alarm probability $\pFA^*$ and at fixed signal strength $\hrms$.
Equivalently\footnote{Due the monotonicity of $\pFD$ as a function of $\hrms$.}
we can fix the false-alarm and false-dismissal probabilities and \emph{minimize}
the required threshold rms signal strength $\hth$, which is the traditional optimization approach
used in previous studies \cite{Brady:1998nj,PhysRevD.72.042004}.

\subsubsection{Gauss approximation for large $\Nseg$}
\label{sec:gauss-appr-large}

One approach (used in \cite{Krishnan:2004sv,PhysRevD.72.042004}) to make further analytical progress
consists in assuming a large number of segments, i.e.\ $\Nseg\gg 1$, and invoke the central limit
theorem to approximate $\chi^2_{4\Nseg}$ by a Gaussian distribution
\begin{equation}
  \label{eq:24}
  \prob{\sumF}{\RHOsumF^2} \stackrel{\Nseg\gg1}{\approx} \left(2\pi \sigma_\sumF^2\right)^{-1/2}
  \exp\left[ - \frac{(\sumF - \av{\sumF})^2}{2\sigma_\sumF^2}\right]\,,
\end{equation}
with mean and variance of $\chi^2_{4\Nseg}(\RHOsumF^2)$ given by
\begin{equation}
  \begin{split}
    \av{\sumF}     &= 4\Nseg + \RHOsumF^2\,,   \\
    \sigma^2_\sumF  &= 2 ( 4\Nseg + 2\RHOsumF^2 )\,.
  \end{split}
\end{equation}
This allows us to analytically integrate Eqs.~(\ref{eq:97a}), (\ref{eq:97b}),
which yields
\begin{align}
  \pFA(\sumF_\thresh) &= \frac{1}{2} \, \erfc\left( \frac{\sumF_\thresh - 4\Nseg}{2\sqrt{4\Nseg}}\right)\,,   \label{eq:25}\\
  \pFD(\sumF_\thresh,\RHOsumF^2) &= \frac{1}{2} \, \erfc\left( \frac{\RHOsumF^2 - (\sumF_\thresh - 4\Nseg)}{2\sqrt{4\Nseg + 2\RHOsumF^2}}\right)\,,   \label{eq:26}
\end{align}
where $\erfc(x) \equiv 1 - \erf(x)$ is the complementary error-function.
Substituting Eq.~(\ref{eq:25}) into Eq.~(\ref{eq:26}), we obtain
\begin{equation}
  \erFD \equiv \frac{\RHOsumF^2 - 2\erFA\sqrt{4\Nseg}}{2 \sqrt{4\Nseg + 2\,\RHOsumF^2}}\,,\label{eq:27}
\end{equation}
where we defined
\begin{equation}
  \label{eq:45}
  \begin{split}
    \erFD &\equiv \erfcInv(2\,\pFD^*) = -\erfcInv(2\,\pdet^*)\,,\\
    \erFA &\equiv \erfcInv(2\,\pFA^*)\,.
  \end{split}
\end{equation}
Solving Eq.~(\ref{eq:27}) for the critical non-centrality parameter $\RHOsumF^{*2}$, we
obtain\footnote{The second solution has $\erFD<0$, corresponding to $\pFD>0.5$.}
\begin{multline}
  \label{eq:82}
  \RHOsumF^{*2}(\erFA,\erFD,\Nseg) = 2\erFA\sqrt{4\Nseg} + 4 \erFD^2\\
  + 2\erFD \sqrt{4\Nseg + 4 \erFA\sqrt{4\Nseg} + 4\erFD^2}\,,
\end{multline}
which we refer to as the ``Gauss approximation''.
Introducing the average per-segment SNR $\RHOF$ as $\RHOF^2 \equiv {\avg{\RHOsumF^2}}/{\Nseg}$, one
can consider two interesting limits of the false-dismissal equation (\ref{eq:27}):
\begin{enumerate}%
\item[(i)] \emph{strong-signal limit} ($\RHOF^2 \gg 1$): the per-segment SNR of the signal is large,
  and we obtain
  \begin{equation}
    \label{eq:23}
    \RHOsumF^* \approx \sqrt{8}\,\erFD\,,
  \end{equation}
  which is somewhat pathological, as $\erFD\gg 1$ and therefore the detection
  probability is \emph{extremely} close to $\pdet=1$. Neither false-alarm threshold nor
  the number of segments $\Nseg$ matter for detectability\footnote{This has been noted previously
    for radio observations\cite{woan_private}} in this case.

\item[(ii)] \emph{weak-signal limit} ($\RHOF^2 \ll 1$): the per-segment SNR of the signal is small,
  and using $\Nseg\gg1$ we find
  \begin{equation}
    \label{eq:27b}
    \RHOsumF^{*2} \approx 2\sqrt{4\Nseg} \, ( \erFA + \erFD )\,,
  \end{equation}
  which we refer to as the ``weak-signal Gauss approximation'' (WSG), which was first used in
  \cite{Krishnan:2004sv} to estimate the sensitivity of the Hough method.
  This approach results in the ``classic'' semi-coherent sensitivity scaling as a function of $\Nseg$, namely
  \begin{equation}
    \label{eq:91}
    {\hth^{-2}}_{,\WSG} =
    \frac{\Ndet}{2\Sn}\,
    \frac{ \left[1 - \pacfactor(\misfmax + \miscmax)\right] } { \erFA + \erFD }\,
    \frac{\Tobs}{\sqrt{\Nseg}}\,.
  \end{equation}
\end{enumerate}

In practice we find that the WSG approximation is often not
well satisfied, and the deviations of the $\Nseg$-scaling in Eq.~(\ref{eq:27b}) from the exact form
of Eq.~(\ref{eq:81}) can lead to dramatically different optimal solutions.
Already the Gauss approximation of Eq.~(\ref{eq:82}) is not well satisfied
for small false-alarm probabilities $\pFA\ll 1$ and segment numbers in the range $\Nseg\lesssim \Ord{1000}$,
as can be seen in Fig.~\ref{fig:SensitivityScalingN}.
A more reliable approximation was recently introduced in \cite{wette2011:_sens}, namely using
the Gaussian distribution only for the false-dismissal equation (\ref{eq:97b}),
while keeping the central $\chi^2$-distribution for the false-alarm equation (\ref{eq:97a}).
For the present work this approach would not be well-suited, however, as we need the
sensitivity equation in the form of a power-law in $\Tobs$ and $\Nseg$, similarly to
Eq.~\eqref{eq:91}.

\subsubsection{Local power-law approximation for $\RHOsumF^*$}
\label{sec:local-power-law}

We can incorporate the exact $\Nseg$-scaling of the critical non-centrality parameter
$\RHOsumF^{*2}$ of Eq.~(\ref{eq:81}) by \emph{locally} expressing it as a power-law in the form
\begin{equation}
  \label{eq:87}
  \RHOsumF^{*2}(\pFA^*,\pFD^*,\,\Nseg) = r_0\, \Nseg^{1 / ( 2\dev)}\,,
\end{equation}
where $\dev(\pFA^*,\pFD^*,\Nseg_0)$ is a parameter quantifying the relative deviation of the exact
$\Nseg$-scaling from the WSG limit of Eq.~\eqref{eq:27b}, where $\dev = 1$.
The power-law coefficients can be computed as
\begin{equation}
  \label{eq:88}
  \dev = \left( 2 \, \frac{\partial \log \RHOsumF^{*2}}{\partial \log\Nseg} \right)^{-1}\,,\quad
  r_0 = \RHOsumF^{*2}\,\Nseg_0^{-1 / ( 2\dev)}\,,
\end{equation}
evaluated at a point $\{\pFA^*,\pFD^*,\Nseg_0\}$.

\begin{figure}[htbp]
  \centering
  \hspace*{-0.05\columnwidth}\includegraphics[width=1.1\columnwidth]{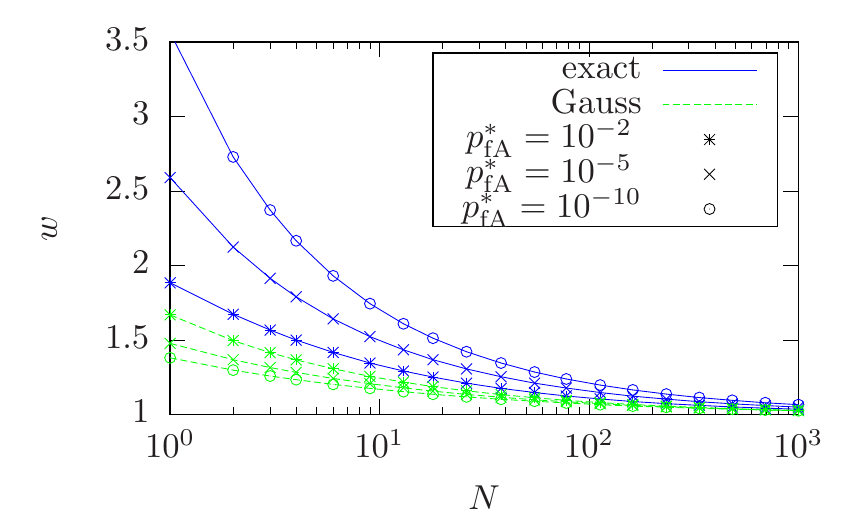}
  \caption{$\Nseg$-scaling coefficient $\dev$ defined in Eq.~\eqref{eq:87} as a function of $\Nseg$,
    for false-dismissal probability $\pFD^*=0.1$, and different false-alarm probabilities
    $\pFA^*\in[10^{-10},10^{-5},10^{-2}]$. Solid lines show the scaling obtained from the exact
    solution Eq.~(\ref{eq:81}), while dashed lines refer to the Gauss approximation
    of Eq.~(\ref{eq:82}). The WSG approximation corresponds to $\dev=1$.
  }
  \label{fig:SensitivityScalingN}
\end{figure}

The function $\dev(\Nseg)$ is shown in Fig.~\ref{fig:SensitivityScalingN}, for a reference
false-dismissal probability of $\pFD^*=0.1$ and different choices of false-alarm probability $\pFA^*$,
both for the exact solution Eq.~(\ref{eq:81}) and for the Gauss approximation of Eq.~(\ref{eq:82}).
We see that the exact $\Nseg$-scaling $\dev$  increasingly deviates from the WSG approximation
($\dev=1$) at \emph{lower} false-alarm probabilities and at smaller $\Nseg$.
The Gauss approximation tends to agree better with the exact scaling at larger $\Nseg$ (as
expected), and at \emph{higher} false-alarm probabilities.

Using the power-law approximation of Eq.~(\ref{eq:87}), we can now express the threshold
signal strength of Eq.~(\ref{eq:99}) as
\begin{align}
    \Lagr_0(\Nseg,\Tobs,\miscmax,\misfmax) &\equiv \frac{r_0\, \Sn}{2\Ndet}\, \hth^{-2} \notag \\
    & = \left[ 1 - \pacfactor ( \miscmax + \misfmax) \right]\,\Tobs \, \Nseg^{-1 / ( 2\dev)}\,,  \label{eq:103}
\end{align}
which defines the objective function $\Lagr_0$ that we want to
maximize as a function of the StackSlide parameters.

We see that, without further constraints the optimal solution would simply be
$\mismax\rightarrow 0$, $\Nseg\rightarrow 1$ and $\Tobs\rightarrow \Tmax$, i.e.\ a fully coherent
search over all the available data $\Tmax$ with an infinitely fine template bank.
This would obviously require infinite computing power, and we therefore need to extend the
optimization problem by a computing-cost constraint.

\subsection{Template counting}
\label{sec:template-counting}

For both the coarse\footnote{We assume a roughly constant number of coarse-grid templates $\Ntc$
  across all segments.} and the fine grid, the respective number of templates\footnote{The
  \emph{templates} in this formulation are not to be confused with the ``patches'' used in BC
  \cite{Brady:1998nj} and CGK \cite{PhysRevD.72.042004}. A ``patch'' in the BC/CGK framework
  corresponds to a \emph{line} of templates along the frequency axis.}
$\{\co{\Nt},\,\ic{\Nt}\}$ covering the parameter space $\dopS$ is given
\cite{2007arXiv0707.0428P,2009PhRvD..79j4017M} by the general expression
\begin{equation}
  \label{eq:9}
  \Nt = \thick_n \, \mismax^{-n/2}\, \vol_n\,,\quad\mbox{with}\quad
  \vol_n \equiv \int_{\dopT_n} \sqrt{\det g}\, d^n\dop\,,
\end{equation}
where $\mismax$ is the maximal-mismatch parameter, $\det g$ is the determinant of the
corresponding parameter-space metric $g_{i j}$,
and $\vol_n$ denotes the metric volume of the $n$-dimensional space $\dopT_n\subseteq\dopS$ spanned
by the template-bank.
The {normalized thickness} $\thick_n$ depends on the geometric structure of the covering, for
example $\thick_{\mathbb{Z}_n} = n^{n/2}\,2^{-n}$ for a hyper-cubic lattice $\mathbb{Z}_n$.

An important subtlety in Eq.~\eqref{eq:9} is the dimensionality $n$ of the
\emph{template-bank space} $\dopT_n$, which can be smaller than the dimensionality of the parameter space
$\dopS$, as previously discussed in \cite{Brady:1998nj,PhysRevD.72.042004}.
The template-bank dimensionality $n$ is generally a (piece-wise constant) function of the StackSlide
parameters $\{\Nseg,\Tseg,\mismax\}$, which determine the metric resolution.
The extent of $\dopS$ along certain directions can be ``thin'' compared to the metric
resolution and would require only a single template along this direction, effectively not
contributing to the template-bank dimensionality.
For different StackSlide parameters, however, the resolution might be sufficient to require more
than one template along this direction, adding to the template-bank dimensionality $n$.

Following \cite{Brady:1997ji,Brady:1998nj,PhysRevD.72.042004}, the correct dimensionality
for given StackSlide parameters can be determined by the condition that $n$ should \emph{maximize}
the number of templates $\Nt_n$ computed via Eq.~(\ref{eq:9}), i.e.\
\begin{equation}
  \label{eq:29}
  \co{\Nt}_{\nc} = \max_{n} \co{\Nt}_n\,,\quad\mbox{and}\quad
  \ic{\Nt}_{\nf} = \max_{n} \ic{\Nt}_n\,.
\end{equation}
This can be understood as follows: if $\Nt$ decreases when adding a template-bank
dimension, then the corresponding parameter-space extent is thinner than the metric
resolution and therefore adds ``fractional'' templates.
On the other hand, if $\Nt$ decreases by removing a dimension, then its extent is thicker than
the metric resolution and requires more than one template to cover it.

An interesting alternative formulation can be obtained by expressing Eq.~\eqref{eq:29} as the
condition $\Nt_n/\Nt_{n-1}>1$ for including an additional dimension $n$.
For constant metrics and simple parameter-space shapes, i.e.\
$\vol_n = \int \sqrt{g}\,d^n\dop = \sqrt{g}\,\Delta\dop_1\times\Delta\dop_2\ldots
\times\Delta\dop_n$, this can be shown to be equivalent to
\begin{equation}
 \label{eq:A5}
 \frac{\thick_n}{\thick_{n-1}}\, \frac{\Delta\lambda_n}{d\lambda_n} >1\,,
\end{equation}
where $d\lambda_n\equiv\sqrt{\mismax\,g^{nn}}$ is the metric \emph{template extent} along dimension
$n$, in terms of the diagonal element $g^{nn}$ of the inverse metric $g^{ij}$.
This shows that Eq.~(\ref{eq:29}) boils down to (apart from the lattice-thickness ratio) the
requirement that the parameter-space extent $\Delta\dop_n$ along a given dimension $n$ must exceed the
corresponding metric template resolution $d\dop_n$.

The coherent (coarse-grid) metric volume $\co{\vol}_\nc$ is typically a steep function of the
coherence time $\Tseg$, and can often be well approximated (over some range of $\Tseg$) by a
power law, namely $\co{\vol}_\nc(\Tseg) \propto \Tseg^{\co{q}}$.
We can therefore write Eq.~(\ref{eq:9}) for $\Ntc$ in the power-law form
\begin{equation}
  \label{eq:119}
  \Ntc_\nc(\miscmax,\Tseg) = \co{k}\,\miscmax^{-\nc/2}\,\Tseg^{\co{q}}\,,
\end{equation}
where $\co{k}=\thick_\nc\co{\vol}_\nc(\Tseg_0)\,\Tseg_0^{-\co{q}}$ for some choice of segment length $\Tseg_0$.

The semi-coherent (fine-grid) metric volume $\ic{\vol}_\nf$ generally depends on both $\Tseg$ and $\Nseg$ and
can typically \cite{Brady:1998nj,Pletsch:2010a,2011arXiv1109.0501M}
be factored in the form
\begin{equation}
  \label{eq:12}
  \ic{\vol}_\nf(\Nseg,\Tseg) = \refac_\nf(\Nseg)\,\co{\vol}_\nf(\Tseg)\,,
\end{equation}
in terms of the \emph{refinement factor} $\refac_\nf(\Nseg) \ge 1$ and the coherent-metric volume
$\co{\vol}_\nf(\Tseg)$ of the fine-grid template space.
Typically $\refac(\Nseg)$ can be well approximated (over some range of $\Nseg$) by a power law,
namely $\refac(\Nseg)\propto \Nseg^{\ic{p}}$.
We can therefore write Eq.~(\ref{eq:9}) for $\Ntf$ in the power-law form
\begin{equation}
  \label{eq:120}
  \Ntf_\nf(\misfmax,\Tseg,\Nseg) = \ic{k}\, \misfmax^{-\nf/2}\,\Tseg^{\ic{q}} \, \Nseg^{\ic{p}} \,,
\end{equation}
where $\ic{k}=\thick_\nf\ic{\vol}_\nf(\Tseg_0,\Nseg_0)\,\Tseg_0^{-\ic{q}}\Nseg_0^{-\ic{p}}$ for some choice of parameters $\{\Tseg_0,\Nseg_0\}$.

\subsection{Computing-cost model}
\label{sec:comp-cost-single}

The total computing cost $\CCtot$ of the interpolating StackSlide statistic
has two main contributions, namely
\begin{equation}
  \label{eq:3}
  \CCtot(\miscmax, \misfmax, \Nseg, \Tseg) = \CCc + \CCf\,,
\end{equation}
where $\CCc(\miscmax,\Nseg,\Tseg)$ is the computing cost of the $\F$-statistic over the coarse grid of
$\co{\Nt}_\nc$ templates for each of the $\Nseg$ segments, and $\CCf(\misfmax,\Nseg,\Tseg)$ is the cost of incoherently
summing these $\F$-values across all segments on a fine grid of $\ic{\Nt}_\nf$ templates.
Note that we neglect all other costs such as data-IO etc, which for any computationally
limited search will typically be much smaller than $\CCtot$.

\subsubsection{Computing cost $\CCc$ of the coherent step}
\label{sec:coher-comp-cost}

The computing cost of the coherent step is
\begin{equation}
  \label{eq:1}
  \CCc(\miscmax,\Nseg,\Tseg) = \Nseg\,\Ntc_\nc(\miscmax,\Tseg)\,\Ndet\,\co{c}_1(\Tseg)\,,
\end{equation}
where $\co{c}_1(\Tseg)$ is the $\F$-statistic computing cost of a single template for a single
segment and a single detector. Here we used the fact that to first order \cite{prix06:_searc} the
number of detectors has no effect on the number of templates $\Ntc$.

As discussed previously in \cite{PhysRevD.72.042004}, there are two fundamentally different
implementations of the $\F$-statistic calculation currently in use: a direct \emph{SFT-method}
\cite{williams99:_effic_match_filter_algor_detec}, and a (generally far more efficient)
\emph{FFT-method} based on barycentric resampling \cite{jks98:_data,2010PhRvD..81h4032P}.
\begin{enumerate}
\item[(i)] The \emph{SFT-method} consists in interpolating frequency bins of short Fourier transforms
  (``SFTs'') of length $T_{\SFT}$, using approximations described in \cite{williams99:_effic_match_filter_algor_detec,prix:_cfsv2}.
  The resulting per-template cost $\co{c}_1(\Tseg)$ is directly proportional to the segment length
  $\Tseg$:
  \begin{equation}
    \label{eq:14}
    \co{c}_1^{\SFT}(\Tseg) = \co{c}_0^{\SFT}\, \frac{\Tseg}{T_\SFT}\,,
  \end{equation}
  where $\co{c}_0^{\SFT}$ is an implementation- and hardware-dependent fundamental computing cost.

\item[(ii)] In the \emph{FFT-method} the cost of searching a frequency band
  $\Delta f$ using an (up-sampled by $u$) FFT frequency-resolution of $u/\Tseg$ is proportional to
  $\Nt_f\,\log 2\Nt_f$, where $\Nt_f = u\Delta f\,\Tseg$ is the number of frequency bins.
  We can therefore express the per-template $\F$-statistic cost $\co{c}_1(\Tseg)$ as
  \begin{equation}
    \label{eq:15}
    \co{c}_1^{\FFT}(\Tseg) = \co{c}_0^{\FFT}\,\log(2u\Delta f\Tseg)\,,
  \end{equation}
  where $\co{c}_0^{\FFT}$ is an implementation- and hardware-dependent fundamental computing cost.
\end{enumerate}

Using the power-law model of Eq.~(\ref{eq:119}) for $\Ntc$,  we can write the coherent computing cost
in the form
\begin{equation}
  \label{eq:47}
  \co{\CC}(\miscmax,\Nseg,\Tseg) = \kappac\, \miscmax^{-\nc/2}\,\Nseg^\etac\, \Tseg^{\deltac}\,,
\end{equation}
where
\begin{equation}
  \label{eq:63}
  \etac = 1,\quad\deltac = \co{q} + \dl\,,
\end{equation}
and where $\dl$ is either
\begin{equation}
  \label{eq:86}
  \dl = \begin{cases}
    \dl_\SFT \equiv 1\,, \\
    \dl_\FFT \equiv \left[\log(2u\Delta f\Tseg_0)\right]^{-1}\,,\\
  \end{cases}
\end{equation}
depending on whether the $\F$-statistic is computed using the \emph{SFT-} or \emph{FFT-method},
respectively.
The expression for $\dl_\FFT$ can be obtained via Eq.~(\ref{eq:33}) and depends (albeit weakly) on
the reference segment length $\Tseg_0$.
The corresponding proportionality factors $\kappac$ are found as
\begin{equation}
  \label{eq:73}
  \begin{split}
  \kappac_{\SFT}  &= \thick_\nc\,\Ndet\,\frac{\co{c}_0^\SFT}{T_\SFT}\,\frac{\co{\vol}_\nc(\Tseg_0)}{\Tseg_0^{\co{q}}}\,,\\
  \kappac_{\FFT}  &=
  \thick_\nc\,\Ndet\,\frac{\co{c}_0^\FFT}{\dl_\FFT}\,\frac{\co{\vol}_\nc(\Tseg_0)}{\Tseg_0^{\deltac}}\,.
  \end{split}
\end{equation}

\subsubsection{Computing cost $\CCf$ of the incoherent step}
\label{sec:incoh-comp-cost}

The computing cost of the StackSlide step is
\begin{equation}
  \label{eq:2}
  \ic{\CC}(\misfmax,\Nseg,\Tseg) = \Nseg\,\ic{\Nt}_\nf(\misfmax,\Tseg,\Nseg)\,\ic{c}_0\,,
\end{equation}
where $\ic{c}_0$ is the implementation- and hardware-dependent fundamental cost of adding one value
of $2\F_k$ for one fine-grid point $\dopf$ in Eq.~(\ref{eq:8}), including the cost of mapping the
fine-grid point to its closest coarse-grid template $\dopc_k(\dopf)$.
The incoherent step operates on coherent multi-detector $\F$-statistic values, and
therefore does not depend on the number of detectors $\Ndet$.

Using the power-law model of Eq.~(\ref{eq:120}) for $\Ntf$, we can write the incoherent
computing cost as
\begin{equation}
  \label{eq:6}
  \CCf(\misfmax,\Nseg,\Tseg) = \ic{\kappa}\, \misfmax^{-\nf/2}\,\Nseg^{\ic{\eta}} \,\Tseg^{\ic{\delta}}\,,
\end{equation}
where
\begin{equation}
  \label{eq:79}
  \etaf = \ic{p}+1\,,\quad \deltaf = \ic{q}\,,
\end{equation}
and the proportionality factor
\begin{equation}
  \label{eq:78}
  \kappaf = \thick_\nf\,\ic{c}_0\,\,
  \frac{ \ic{\vol}_\nf(\Nseg_0,\Tseg_0) }{ \Nseg_0^{\ic{p}}\,\Tseg_0^{\ic{q}}}\,,
\end{equation}
for given reference values $\{\Nseg_0,\Tseg_0\}$.

\subsubsection{General power-law computing-cost model}

Combining Eqs.~(\ref{eq:47}) and (\ref{eq:6}) we arrive at the following power-law model for the
total computing cost, defined in Eq.~(\ref{eq:3}), namely
\begin{equation}
  \label{eq:18}
    \CCtot =
    \kappac\,\miscmax^{-\nc/2}\,\Nseg^\etac\,\Tseg^\deltac
    + \kappaf\,\misfmax^{-\nf/2}\,\Nseg^\etaf\, \Tseg^\deltaf\,.
\end{equation}

If a given computing-cost function does not follow this model, we can always produce a local fit to
Eq.~\eqref{eq:18}, which should be valid over some range of parameters $\{\Tseg,\Nseg\}$, namely
\begin{equation}
  \delta \equiv \left. \frac{\partial \log \CC}{\partial \log\Tseg}\right.\,,\quad
  \eta   \equiv \left. \frac{\partial \log \CC}{\partial \log\Nseg}\right.\,,  \label{eq:33}
\end{equation}
\begin{equation}
  \label{eq:121}
  \kappa \equiv \frac{\CC(\mismax_0,\Nseg_0,\Tseg_0)}{\mismax_0^{-n/2}\,\Nseg_0^{\eta}\,\Tseg_0^{\delta}}\,,
\end{equation}
for reference values $\{\mismax_0,\Nseg_0,\Tseg_0\}$.
Note that $\delta$ generally depends only on $\Tseg_0$, while $\eta$ depends only on $\Nseg_0$,
due to the way these dependencies typically factor (cf.\ Sec.~\ref{sec:comp-cost-single}).
The mismatch dependency $\propto \mismax^{-n/2}$ is exact according to Eq.~(\ref{eq:9}), but a given
computing-cost function might still deviate from this behaviour
(e.g.\ the BC/CGK computing-cost function discussed in Sec.~\ref{sec:comp-prev-stud}).
In this case one can extend the power-law fit by extracting the ``mismatch-dimension'' $n$ via
\begin{equation}
  \label{eq:38}
  n \equiv - 2 \, \left. \frac{\partial \log \CC}{\partial \log\mismax}\right.\,.
\end{equation}

It will be more convenient in the following to work in terms of $\{\Nseg,\Tobs\}$ instead of
$\{\Nseg,\Tseg\}$, where $\Tobs = \Nseg\,\Tseg$ is the total time span of data used.
Changing variables, we obtain the computing-cost model in the form
\begin{equation}
  \label{eq:17}
    \CCtot =
    \kappac\,\miscmax^{-\nc/2}\,\Nseg^{-\epsc}\,\Tobs^\deltac
     + \kappaf\,\misfmax^{-\nf/2}\,\Nseg^{-\epsf}\, \Tobs^\deltaf\,,
\end{equation}
where we defined
\begin{equation}
  \label{eq:32}
  \eps \equiv \delta - \eta\,,
\end{equation}
generally satisfying $\eps > 0$ for all realistic cases considered here.
Note that $\mismax$ and $\Nseg$ are dimensionless, therefore the respective units of $[\CC/\kappa]$
are $[\Tobs^\delta]$.

\begin{table*}[htbp]
  \centering
  \begin{tabular}{c || l | l | l}
    Symbol   	& {\hfill Description\hfill}     			& {\hfill Relations\hfill\,} 	&  {\hfill Refs\hfill\,}\\\hline
    $\Nseg$  	& Number of segments	   				& 				& Sec.~\ref{sec:prop-single-stage}, Eq.~\eqref{eq:7}   \\
    $\Tseg$  	& Segment duration					& 				& Sec.~\ref{sec:prop-single-stage} \\
    $\Tobs$  	& Total observation time				& $\Tobs = \Nseg \, \Tseg$ 	& Sec.~\ref{sec:prop-single-stage} \\
    $\co{Q}$ 	& a quantity $Q$ referring to the \emph{coherent} step	& 				& Sec.~\ref{sec:prop-single-stage} \\
    $\ic{Q}$ 	& a quantity $Q$ referring to the \emph{incoherent} step& 				& Sec.~\ref{sec:prop-single-stage} \\
    $n$	     	& number of template-bank dimensions			& 				& Eq.~\eqref{eq:29}   \\
    $\mismax$ 	& maximal template-bank mismatch parameter		& 				& Eqs.~\eqref{eq:114},~\eqref{eq:115}\\
    $\pacfactor$& average mismatch factor $\in[0,1]$			& $\avg{\mis} = \pacfactor\, \mismax$				&  Eq.~\eqref{eq:39} \\
    $\dev$      & $\Lagr_0$ sensitivity scaling with $\Nseg$		& $\left.\Lagr_0\right|_{\mismax,\Tobs} \propto \Nseg^{-1/(2\dev)}$	&  Eq.~\eqref{eq:87} \\
    $\varpi$	& Lagrange multiplier for computing-cost constraint	& $\Lagr = \Lagr_0 + \varpi(\CCtot-\CC_0)$				&  Eq.~\eqref{eq:66} \\
    $\kappa$ 	& computing-cost prefactor	 			& 								&  Eqs.~\eqref{eq:47},\eqref{eq:6} \\
    $\delta$ 	& computing-cost $\Tobs$- or $\Tseg$- exponent at fixed $\Nseg$& $\left.\CC\right|_{\Nseg} \propto \Tseg^\delta \propto \Tobs^\delta$  &  Eqs.~\eqref{eq:47},\eqref{eq:6}\\
    $\eta$   	& computing-cost $\Nseg$-exponent at fixed $\Tseg$  	& $\left.\CC\right|_{\Tseg} \propto \Nseg^\eta$ 			&  Eqs.~\eqref{eq:47},\eqref{eq:6}\\
    $-\eps$   	& computing-cost $\Nseg$-exponent at fixed $\Tobs$  	& $\left.\CC\right|_{\Tobs} \propto \Nseg^{-\eps}$ 		&  Eq.~\eqref{eq:17} \\
  \end{tabular}
  \caption{Overview of symbols and notation used in the formulation of the optimization problem.}
  \label{tab:symbols}
\end{table*}

\section{Maximizing  sensitivity at fixed computing cost}
\label{sec:optim-sens-stacksl}

We want to maximize the objective function $\Lagr_0\propto\hth^{-2}$ defined in Eq.~(\ref{eq:103})
under the constraint of fixed computing cost, $\CCtot = \CC_0$.
We therefore need to find the stationary points of the Lagrange function
\begin{equation}
  \label{eq:66}
  \Lagr(\Nseg,\Tobs,\miscmax,\misfmax,\varpi) = \Lagr_0 + \varpi \, [ \co{\CC} + \ic{\CC} - \CC_0]\,,
\end{equation}
where stationarity with respect to the Lagrange multiplier, i.e.\ $\partial_\varpi \Lagr = 0$,
returns the computing-cost constraint $\co{\CC} + \ic{\CC} = \CC_0$.

Table~\ref{tab:symbols} provides a ``dictionary'' summarizing the notation used here and in the
previous section to formulate the optimization problem.

Before embarking on the full optimization problem, it is instructive to consider two special cases,
namely (i) a fully coherent search, and
(ii) searches where the computing cost is dominated by one contribution, either coherent $\CCc$ or
incoherent $\CCf$.

\subsection{Special case (i): Fully coherent search}
\label{sec:fully-coher-search}

The fully coherent search is a special case of Eq.~(\ref{eq:66}) with the additional constraint
$\Nseg=1$, and therefore \ $\Tseg = \Tobs$, $\misfmax=0$, and $\ic{\CC}= 0$.
This leaves us with the reduced Lagrangian
\begin{equation}
  \label{eq:35}
  \Lagr(\Tobs,\miscmax,\varpi) = ( 1 - \pacfactor \miscmax)\,\Tobs
  + \varpi \, [   \kappac\,\miscmax^{-\nc/2}\,\Tobs^\deltac - \CC_0 ]\,.
\end{equation}
Requiring stationarity with respect to $\{\Tobs,\miscmax,\varpi\}$ results in the optimal
StackSlide parameters
\begin{align}
  \pacfactor\, \opt{\miscmax} &= \left( 1 + \frac{2\deltac}{\nc}\right)^{-1}\,, \label{eq:53}\\
  \opt{\Tobs} &= \left(\frac{\CC_0}{\kappac}\right)^{1/\deltac}\,\opt{\miscmax}^{\nc/(2\deltac)}\,. \label{eq:53b}
\end{align}
Interestingly the optimal mismatch $\opt{\miscmax}$ is independent of both the computing-cost
constraint $\CC_0$ and the observation time $\Tobs$.
The scaling of the resulting threshold signal strength $\hth$ with computing cost $\CC_0$ is
therefore
\begin{equation}
  \hth^{-1} \propto \CC_0^{1/(2\deltac)}\,.\label{eq:53c}
\end{equation}
In practical applications we often find $\deltac \approx 3-7$, and so $\opt{\Tobs}$ and $\hth^{-1}$
will increase very slowly with increasing computing cost $\CC_0$. This indicates that a brute-force
approach of throwing more computing power at a fully coherent search will typically yield meagre
returns in sensitivity.

\subsection{Special case (ii): Computing cost dominated by one contribution}
\label{sec:comp-cost-domin}

If either the coherent $\CCc$ or incoherent $\CCf$ contribution dominates the total computing
cost (\ref{eq:17}), we can write
\begin{equation}
  \label{eq:55}
  \CCtot \approx \kappa\,\mismax^{-n/2}\,\Nseg^{-\eps}\,\Tobs^\delta\,,
\end{equation}
where all StackSlide parameters now refer to dominant contribution only.

We assume that the negligible computing-cost contribution implies that we can also neglect the
corresponding mismatch: if the respective step is cheap, one can easily increase sensitivity by
reducing the corresponding mismatch until it is negligible, i.e.\ we assume
$\avg{\mis_\semi} \approx \pacfactor\,\mismax$. This qualitative argument will be confirmed by the
general solution in the next section.
We can therefore write the objective function Eq.~\eqref{eq:103} as
\begin{equation}
  \label{eq:56}
  \Lagr_0(\Nseg,\Tobs,{\mismax} ) \approx (1 - \pacfactor\,\mismax)\,\Nseg^{-1/(2\dev)}\,\Tobs\,.
\end{equation}
Using Eq.~(\ref{eq:55}) we can obtain
\begin{align}
  \Nseg(\CC_0,\mismax,\Tobs) &= (\CC_0/\kappa)^{-1/\eps}\,\mismax^{-n/(2\eps)}\,\Tobs^{\delta/\eps}\,,  \label{eq:58}\\
  \Tseg(\CC_0,\mismax,\Tobs) &= (\CC_0/\kappa)^{1/\eps}\,\mismax^{n/(2\eps)}\,\Tobs^{-\eta/\eps}\,,  \label{eq:4}
\end{align}
which shows that increasing $\Tobs$ at fixed $\CC_0$ results in more and shorter segments, while
increasing $\CC_0$ at fixed $\Tobs$ results in fewer and longer segments (assuming $\eps>0$).
Substituting this into Eq.~(\ref{eq:56}) yields the threshold signal strength
\begin{equation}
  \label{eq:5}
\hth^{-2}\propto (\CC_0/\kappa)^{1/(2\dev\eps)}\,
\left[(1-\pacfactor\mismax)\,\mismax^{n/(4\dev\eps)}\right]\,\,\Tobs^{\,\crit/{(2\dev\eps)}}\,,
\end{equation}
where we introduced the parameter
\begin{equation}
  \label{eq:104}
  \crit \equiv 2\dev\eps - \delta\,,
\end{equation}
which will be of critical importance in determining the character of the optimal solution.

The objective function $\Lagr_0\propto \hth^{-2}$ can be easily maximized over mismatch $\mismax$, resulting in
\begin{equation}
  \label{eq:59}
  \pacfactor\,\optO{\mismax} = \left[ 1 + \frac{4\dev\eps}{n}\right]^{-1}\,,
\end{equation}
which is independent of both $\CC_0$ and $\Tobs$. This solution differs from
Eq.~(\ref{eq:53}) of the fully coherent case, even when the coherent cost dominates (where $\epsc = \deltac-1$).

We see in Eq.~(\ref{eq:5}) that there is \emph{no extremum} of $\hth$ (at least in regions of
approximately constant power-law exponents).
Given that $\dev\ge1$ and generally $\eps>0$, we can distinguish two different regimes
depending on the sign of critical scaling exponent $\crit$ defined in Eq.~\eqref{eq:104}:
\begin{description}

\item[$\crit>0$:] sensitivity improves (i.e.\ $\hth^{-1}$ increases) with \emph{increasing} $\Tobs$ (at fixed $\CC_0$).
  Therefore sensitivity is only limited by the total amount of data $\Tmax$ available.

\item[$\crit < 0$:] sensitivity improves with
  \emph{decreasing} $\Tobs$, so one should use less data (until the assumptions change).

\end{description}
In practice these extreme conclusions will be modified, as the power-law exponents
will vary (slowly) as functions of $\Nseg$ and $\Tobs$, and the assumption of a dominating
computing-cost contribution might also no longer be satisfied.
The marginal case $\crit = 0$ marks a possible sensitivity maximum, namely if increasing
$\Tobs$ results in $\crit <0$ and decreasing $\Tobs$ leads to $\crit > 0$.

We can obtain a useful qualitative picture of the full optimization problem by considering the
two extreme cases of dominating computing contribution $\CCc$ or $\CCf$:
\begin{itemize}
\item if $\CCc\gg\CCf$: we always have $\critc > 0$ (for all cases of interest  $\etac=1$,
  $\deltac > 2$ and $\dev\ge 1$). Therefore sensitivity improves with increasing $\Tobs$.
  As seen in Sec.~\ref{sec:monotony-laws} this shifts computing cost to the
  incoherent contribution. Eventually one either uses all the data $\Tmax$ or the coherent cost no
  longer dominates.

\item if $\CCf\gg\CCc$: the incoherent parameter $\critf$ can have any sign.
  If $\critf > 0$ one would \emph{increase} $\Tobs$ until all the data $\Tmax$ is used (or we reach $\critf=0$).
  If $\critf < 0$ one would \emph{decrease} $\Tobs$ until the incoherent cost no longer dominates.
\end{itemize}
These limiting cases show that the \emph{type} of optimal solution will be determined
solely by the \emph{incoherent} critical exponent $\critf =2\dev\epsf - \deltaf$, namely

\begin{equation}
  \label{eq:42}
  \optO{\Tobs} = \begin{cases}
    \text{finite} & \text{if $\critf \le 0$}\,,\\
    \infty        & \text{otherwise}\,,
  \end{cases}
\end{equation}
which we refer to as the \emph{bounded} and the \emph{unbounded} regime, respectively.

\subsection{General optimality conditions}
\label{sec:analyt-solut-gener}

We now return to the full optimization problem of Eq.~(\ref{eq:66}), namely
\begin{equation}
  \Lagr = \Lagr_0 + \varpi \, [ \co{\CC} + \ic{\CC} - \CC_0]\,,   \label{eq:104a}
\end{equation}
where
\begin{align}
  \Lagr_0 &= \left[ 1 - \pacfactor ( \miscmax + \misfmax) \right]\,{\Tobs}\,\Nseg^{-1/(2\dev)}\,,\label{eq:104b}\\
  \co{\CC} &= \kappac\,\miscmax^{-\nc/2}\,\Nseg^{-\epsc}\,\Tobs^\deltac\,,\label{eq:104c}\\
  \ic{\CC} &= \kappaf\,\misfmax^{-\nf/2}\,\Nseg^{-\epsf}\, \Tobs^\deltaf\,.\label{eq:104d}
\end{align}
It will be useful introduce the computing-cost \emph{ratio}
\begin{equation}
  \label{eq:50}
  \cratio \equiv {\CCc}/{\CCf}\,,
\end{equation}
and express the respective contributions as
\begin{equation}
  \label{eq:62}
  \co{\CC} = \frac{\CC_0}{1 + \cratio^{-1}}\,,\quad
  \ic{\CC} = \frac{\CC_0}{1 + \cratio}\,.
\end{equation}
Using Eqs.~(\ref{eq:104c}), (\ref{eq:104d}) to solve for $\Tobs$ and $\Nseg$, respectively, we
obtain
\begin{align}
  \Nseg^\detCoefs &= \frac{(\CC_0/\kappaf)^{\deltac}}{(\CC_0/\kappac)^{\deltaf}}\,\frac
  {\left[ \miscmax^{-\nc/2} \,(1+\cratio^{-1})\right]^\deltaf}
  {\left[ \misfmax^{-\nf/2} \,(1+\cratio) \right]^\deltac}\,,  \label{eq:100a}\\
  \Tobs^\detCoefs &= \frac{ (\CC_0/\kappaf)^{\epsc}}{ (\CC_0/\kappac)^\epsf}\, \frac
  {\left[ \miscmax^{-\nc/2} \,(1+\cratio^{-1}) \right]^\epsf }
  {\left[ \misfmax^{-\nf/2} \,(1+\cratio)     \right]^\epsc }\,, \label{eq:100b}
\end{align}
where $\detCoefs$ is the determinant of the matrix
$[\deltac, \etac;\deltaf,\etaf]$, which for all cases of practical interest seems to be positive
definite, namely
\begin{equation}
  \label{eq:60}
  \detCoefs \equiv \deltac\etaf - \deltaf\,\etac > 0\,.
\end{equation}
The segment length $\Tseg=\Tobs/\Nseg$ can similarly be obtained as
\begin{equation}
  \label{eq:16}
  \Tseg^\detCoefs = \frac{(\CC_0/\kappac)^{\etaf}}{(\CC_0/\kappaf)^{\etac}}\,\frac
  {\left[ \misfmax^{-\nf/2} \,(1+\cratio) \right]^\etac}
  {\left[ \miscmax^{-\nc/2} \,(1+\cratio^{-1})\right]^\etaf}
  \,.
\end{equation}

\subsubsection{Stationarity with respect to mismatches $\{\miscmax,\misfmax\}$}
\label{sec:vari-with-resp}

Requiring stationarity with respect to the mismatches, i.e.\
$\partial_{\miscmax} \Lagr = \partial_{\misfmax} \Lagr = 0$, yields
\begin{equation}
  \label{eq:34}
  \begin{split}
    \varpi \co{\CC} &= - 2 \pacfactor\,\frac{\opt{\miscmax}}{\nc}\,{\Tobs}\Nseg^{-1/(2\dev)}\,,\\
    \varpi \ic{\CC} &= - 2 \pacfactor\,\frac{\opt{\misfmax}}{\nf}\,{\Tobs}\Nseg^{-1/(2\dev)}\,,
  \end{split}
\end{equation}
which results in the remarkable relation
\begin{equation}
  \label{eq:65}
  \frac{\opt{\miscmax}/\nc}{\opt{\misfmax}/\nf} = \cratio\,.
\end{equation}
The ratio of optimal mismatch per dimension is simply given by the computing-cost ratio $\cratio$.
This result confirms an assumption made in Sec.~\ref{sec:comp-cost-domin} about the optimal
solution, namely that a negligible computing-cost contribution also implies that one can neglect
the corresponding mismatch.

\subsubsection{Stationarity with respect to number of segments $\Nseg$}
\label{sec:vari-with-resp-1}

Requiring stationarity with respect to $\Nseg$ (treated as continuous), i.e. $\partial_\Nseg \Lagr=0$ yields
\begin{equation}
  \label{eq:67}
  \Lagr_0 + 2\dev\,\varpi\left[ \epsc\,\co{\CC} + \epsf\,\ic{\CC}\right] = 0\,,
\end{equation}
and substituting Eqs.~(\ref{eq:34}) and (\ref{eq:104b}), we obtain
\begin{equation}
  \label{eq:36}
  \frac{\opt{\miscmax}}{\optO{\miscmax}} + \frac{\opt{\misfmax}}{\optO{\misfmax}} = 1\,,
\end{equation}
where we used the asymptotic optimal mismatches $\optO{\mismax}$ defined in Eq.~(\ref{eq:59}) for
the two limiting cases of dominating coherent or incoherent computing-cost, respectively.
Equation \eqref{eq:36} can be interpreted as defining a two-dimensional ellipse in $\sqrt{\mismax}$
with semi-major axes $\sqrt{\optO{\mismax}}$. Combining this with Eq.~(\ref{eq:65}) we obtain the
optimal mismatches
\begin{equation}
  \label{eq:48}
  \begin{split}
    \frac{\nc}{\opt{\miscmax}} &= \frac{\nc}{\optO{\miscmax}} + \frac{\nf}{\optO{\misfmax}}\, \cratio^{-1}\,,\\
    \frac{\nf}{\opt{\misfmax}} &= \frac{\nf}{\optO{\misfmax}} + \frac{\nc}{\optO{\miscmax}}\, \cratio \,,
  \end{split}
\end{equation}
which reduces to the limiting cases of Eq.~(\ref{eq:59}) when either computing cost dominates, i.e.\
when $\cratio \ll 1$ or $\cratio \gg 1$.
We can express the optimal mismatch prefactor in Eq.~(\ref{eq:104b}) as
\begin{equation}
  \label{eq:69}
  \opt{\left[ 1 - \pacfactor(\miscmax+\misfmax)\right]} =
  \left[ 1 + \frac{1}{4\dev}\frac{ \co{n}\cratio + \ic{n} }{\co{\eps}\cratio + \ic{\eps}}\right]^{-1}\,.
\end{equation}

The optimal mismatches Eq.~(\ref{eq:48}) only depend on the computing-cost ratio $\cratio$.
Substituting into Eq.~(\ref{eq:100b}) we therefore obtain a relation of the form
$\Tobs_0 = \Tobs(\CC_0,\opt{\cratio})$ for given observation time $\Tobs_0$, which can (numerically) be solved for
$\opt{\cratio} = \cratio(\CC_0,\Tobs_0)$.
Similarly, one could specify $\Nseg_0$ and solve Eq.~(\ref{eq:100a}) for
$\opt{\cratio} = \cratio(\CC_0,\Nseg_0)$.
In either case the optimal mismatches are obtained from Eq.~(\ref{eq:48}) and the optimal number
and length of segments from Eqs.~(\ref{eq:100a}) and (\ref{eq:100b}), fully closing the optimal solution at fixed $\Tobs$.

\subsubsection{Monotonicity relations with $\Tobs$}
\label{sec:monotony-laws}

It is interesting to consider the behaviour of the optimal ``fixed-$\Tobs$'' solution of the previous
section as a function of $\Tobs$.
We see in Eq.~(\ref{eq:48}) that $\opt{\miscmax}$ is monotonically increasing with $\cratio$, while
$\opt{\misfmax}$ is decreasing, i.e.\
\begin{equation}
  \label{eq:21}
  \partial_\cratio \opt{\miscmax}>0,\quad\mbox{and}\quad \partial_\cratio\opt{\misfmax}<0\,.
\end{equation}
We generally assume $\detCoefs \equiv \deltac \etaf - \deltaf\etac > 0$ and $\eps>0$ which implies
that the right-hand side of Eq.~(\ref{eq:100b}) is monotonically \emph{decreasing} with
$\cratio$, while the left-hand side is monotonically increasing with $\Tobs$.
Therefore $\cratio$ must be montonically \emph{decreasing} with $\Tobs$, i.e.\
\begin{equation}
  \label{eq:22}
  \partial_\Tobs \cratio < 0\,.
\end{equation}
Therefore the optimal solution shifts computing cost from the coherent to the incoherent step
with increasing $\Tobs$, which had already been used in Sec.~\ref{sec:comp-cost-domin}.
Combining this with Eq.~(\ref{eq:21}) we find
\begin{equation}
  \label{eq:30}
  \partial_\Tobs \opt{\miscmax} < 0,\quad\mbox{and}\quad \partial_\Tobs\opt{\misfmax} >0\,,
\end{equation}
and using this with Eqs.~(\ref{eq:100a}) and (\ref{eq:16}), we can further deduce
\begin{equation}
  \label{eq:31}
  \partial_\Tobs  \opt{\Nseg} > 0\,,\quad\mbox{and}\quad
  \partial_\Tobs  \opt{\Tseg} < 0\,,
\end{equation}
namely increasing $\Tobs$ results in more segments of shorter duration.

\subsubsection{Stationarity with respect to observation time $\Tobs$}
\label{sec:optimize-at-variable}

Requiring stationarity of $\Lagr$ with respect to $\Tobs$, i.e.\ $\partial_\Tobs
\Lagr = 0$, yields the final condition
\begin{equation}
  \label{eq:74}
  \Lagr_0 + \varpi\left[ \deltac\,\co{\CC} + \deltaf\,\ic{\CC}\right] = 0\,,
\end{equation}
which combined with Eq.~(\ref{eq:67}) results in
\begin{equation}
  \label{eq:75}
  \critc\,\CCc + \critf\,\CCf = 0\,,
\end{equation}
where the critical exponents $\crit$ are defined in Eq.~(\ref{eq:104}).
We generally expect $\critc > 0$, as discussed in Sec.~\ref{sec:comp-cost-domin}, and therefore the
stationarity condition can only have a solution if
\begin{equation}
  \label{eq:101}
  \critf \equiv \deltaf - 2\etaf + 2(\dev-1)\,\epsf \,<\, 0\,.
\end{equation}
This conclusion is consistent with the analysis of Sec.~\ref{sec:comp-cost-domin}:
$\critf < 0$ characterizes a \emph{bounded} regime with finite optimal $\optO{\Tobs}$, while
$\critf>0$ characterizes an \emph{unbounded} regime with $\optO{\Tobs}\rightarrow\infty$.

If $\optO{\Tobs}$ exceeds the available data $\Tmax$, then we simply apply the fixed-$\Tobs$ solution of
Sec.~\ref{sec:vari-with-resp-1}.
Otherwise Eq.~(\ref{eq:75}) directly yields
\begin{equation}
  \label{eq:76}
  \opt{\cratio} = -\frac{\critf}{\critc}\,,
\end{equation}
closing the optimal solution via Eqs.~(\ref{eq:48}), (\ref{eq:100a}) and (\ref{eq:100b}).

\subsubsection{Monotonicity relations with $\CC_0$}
\label{sec:monot-relat-with-C0}

For a bounded optimal solution with $\optO{\Tobs}\le\Tmax$, we see from
Eq.~(\ref{eq:76}) that $\opt{\cratio}$ and $\{\opt{\miscmax},\,\opt{\misfmax}\}$ are independent of
the computing-cost constraint $\CC_0$.
Inserting Eqs.~(\ref{eq:100a}),(\ref{eq:100b}) into Eq.~(\ref{eq:103}), we can therefore
read off the scaling
\begin{equation}
  \label{eq:49}
  \hth^{-1} \propto \CC_0^{(\critc - \critf)/(4\dev\detCoefs)}\,,
\end{equation}
which shows that any ``reasonable'' search should satisfy
\begin{equation}
  \label{eq:61}
  \critc > \critf\,,
\end{equation}
in order for sensitivity to \emph{improve} with increasing $\CC_0$ (assuming $\detCoefs>0$).
Furthermore, from Eqs.~(\ref{eq:100a}), (\ref{eq:100b}) and
(\ref{eq:16}) we obtain the monotonicity relations:
\begin{equation}
  \label{eq:80}
  \begin{split}
    \partial_{\CC_0} \opt{\Nseg} &\propto \deltac - \deltaf\,,\\
    \partial_{\CC_0} \opt{\Tobs} &\propto \epsc - \epsf\,,\\
    \partial_{\CC_0} \opt{\Tseg} &\propto \etaf - \etac\,.
  \end{split}
\end{equation}
We expect $\etaf > \etac = 1$, therefore the optimal segment length $\opt{\Tseg}$ will generally
increase with $\CC_0$.

The behaviour of the optimal number of segments is less clear-cut:
if $\deltac < \deltaf$ then $\opt{\Nseg}$ \emph{decreases} with $\CC_0$, which can result in a fully
coherent search being optimal, despite $\optO{\Tobs} < \Tmax$.
A StackSlide search is therefore not guaranteed to be more sensitive than a fully coherent
search at the same computing power, even when computationally limited.

Similarly, $\opt{\Tobs}$ can either increase with $\CC_0$ (if $\epsc>\epsf$), or decrease:
a more expensive and more sensitive search can be using less data.

\section{Examples of practical application}
\label{sec:exampl-appl}

In order to illustrate the practical application of this analytical framework and its potential
gains in sensitivity we consider a few different examples of CW searches.

\subsection{Directed searches for isolated neutron stars}
\label{casa}

Directed searches target NSs with \emph{known}
sky-position but unknown frequency and frequency derivatives, i.e. $\{f, \dot{f}, \ddot{f}, \ldots\}$.
The approximate phase metric of this parameter space for isolated NSs is known analytically and
constant over the parameter space, e.g.\ see [Eq.~(10) in \cite{2008CQGra..25w5011W}].
The number of coarse-grid templates scales as
\begin{equation}
  \label{eq:41}
  \co{\Nt} \propto \Tseg^{n(n+1)/2}\,,
\end{equation}
while the refinement of the semi-coherent metric [Eq.~(92) in \cite{Pletsch:2010a}] scales as
\begin{equation}
  \label{eq:64}
  \gamma_n \propto \Nseg^{n(n-1)/2}\,.
\end{equation}
The coherent computing-cost exponents Eq.~(\ref{eq:63}) are therefore
\begin{equation}
  \label{eq:83}
  \deltac = \nc(\nc+1)/2 + \dl\,,\quad\etac=1\,,
\end{equation}
where $\dl$ depends on the $\F$-statistic implementation as given by Eq.~(\ref{eq:86}).
The incoherent computing-cost exponents Eq.~(\ref{eq:79}) are
\begin{equation}
  \label{eq:68}
  \deltaf = \nf(\nf+1)/2\,,\quad \etaf = 1+\nf(\nf-1)/2\,,
\end{equation}
which results in $\epsf = \nf-1$.

For $\nc\ge2$ the condition $\critc = \deltac-2+2(\dev-1)\epsc > 0$ holds as expected, while the
critical boundedness parameter of Eq.~(\ref{eq:101}) now reads as
\begin{equation}
  \label{eq:70}
  \critf = \frac{\nf}{2}( 3 - \nf ) - 2 +2(\dev-1)(\nf-1)\,,
\end{equation}
which for the first few values of $n$ evaluates to
\begin{equation}
  \label{eq:37}
  \begin{split}
    \critf_1 &= -1\,,\\
    \critf_2 &= -3 + 2\dev\,,\\
    \critf_3 &= -6 + 4\dev\,,\\
    \critf_4 &= -10 + 6\dev.
  \end{split}
\end{equation}
In the WSG limit (i.e.\ $\dev\rightarrow 1$) this is always $\critf <0$, and therefore the search
falls into the bounded regime. However, in general $\dev>1$ and therefore directed StackSlide
searches can be either bounded or unbounded.

\subsubsection*{Directed search for Cassiopeia-A}
\label{sec:search-cassiopeia}

As a concrete example we consider the directed search for the compact object in Cassiopeia-A (CasA).
This search has been performed using LIGO S5 data, and the
resulting upper limits have been published in \cite{2010ApJ...722.1504A}.
For the present example we use the search setup as originally proposed in \cite{2008CQGra..25w5011W},
namely a fully coherent $\F$-statistic search (using the ``SFT'' method, i.e.\ $\dl=1$) using data
spanning $\Tobs = 12\,\days$, with a maximal template-bank mismatch of $\miscmax = 0.2$.
The setup assumed two detectors with identical noise floor $\Sn$ and a $70\%$ duty cycle, which
we can formally incorporate as $\Ndet= 2 \times 0.7 = 1.4$ in Eqs.~(\ref{eq:1}) and (\ref{eq:99}).
The parameter space spanned a frequency range of $f\in[100,300]\,\Hz$ and
spindown-ranges corresponding to a spindown-age of $\tau=300\,\years$, see \cite{2008CQGra..25w5011W}.
The template-bank dimension for the given StackSlide parameters was determined as $\nc=3$,
resulting in a power-law scaling of $\deltac = 7$ according to Eq.~(\ref{eq:83}).

In order to compare sensitivity estimates of different search setups, we use nominal (per-template)
false-alarm and false-dismissal probabilities of
\begin{equation}
  \label{eq:40}
  \pFA^*=10^{-10}, \qquad \pFD^* = 0.1\,.
\end{equation}
We use a rough estimate of $\pacfactor = 0.5$ (e.g.\ see [Fig.~8 in \cite{2009PhRvD..79j4017M}]) for
the geometric average-mismatch factor of the $A_3^*$-lattice that was used in this search.
Integrating Eqs.~(\ref{eq:97a}),(\ref{eq:97b}) and solving for $\RHOF^*$ yields $\RHOF^* \approx 8.35$.
Substituting this into Eq.~(\ref{eq:99}) with $\misfmax=0$, $\miscmax=0.2$ yields an estimate for
the weakest detectable signal $\hth$ of the original $\F$-statistic search:
\begin{equation}
  \label{eq:97}
  \left. \frac{\hth}{\sqrt{\Sn}} \right|_{\mathrm{ref}} \approx 5.2\times 10^{-3}\,\sqrt{\Hz},
\end{equation}

Timing a current StackSlide code using the {SFT-method}, one can extract
approximate timing parameters
\begin{equation}
  \label{eq:19}
  \co{c}_0^\SFT = 7\times 10^{-8}\,\secs,\qquad \ic{c}_0 = 6\times 10^{-9}\,\secs\,,
\end{equation}
which results in a total computing cost for the original search\footnote{Using the original timing
  constant $\co{c}_\SFT^{(0)}=6\times10^{-7}\,\secs$ of \cite{2008CQGra..25w5011W}, we correctly
  recover the original estimate of $\co{\CC}\approx 20\times200\,\days$} of $\CC_0 \approx 472\,\days$
on a single cluster node. This number is used as the computing-cost constraint
$\CC_0$ for this example.

First we consider an optimal \emph{coherent} search as described in Sec.~\ref{sec:fully-coher-search},
namely using Eq.~(\ref{eq:53}) we find
\begin{equation}
  \label{eq:71}
  \pacfactor\, \opt{\miscmax} \approx 0.18\quad \Longrightarrow \opt{\miscmax}\approx 0.36\,,
\end{equation}
and using Eq.~(\ref{eq:53b}) this results in $\opt{\Tobs}=13.6\,\days$, which is only
about $\sim13\%$ longer then the original search proposal of \cite{2008CQGra..25w5011W}.
The total improvement in the minimal signal strength $\hth$ is less than $2\%$ compared to
Eq.~(\ref{eq:97}), which shows that the original search proposal was remarkably close to an optimal
coherent search.

\begin{figure*}[htbp]
\centering
\mbox{
  \includegraphics[width=0.8\columnwidth]{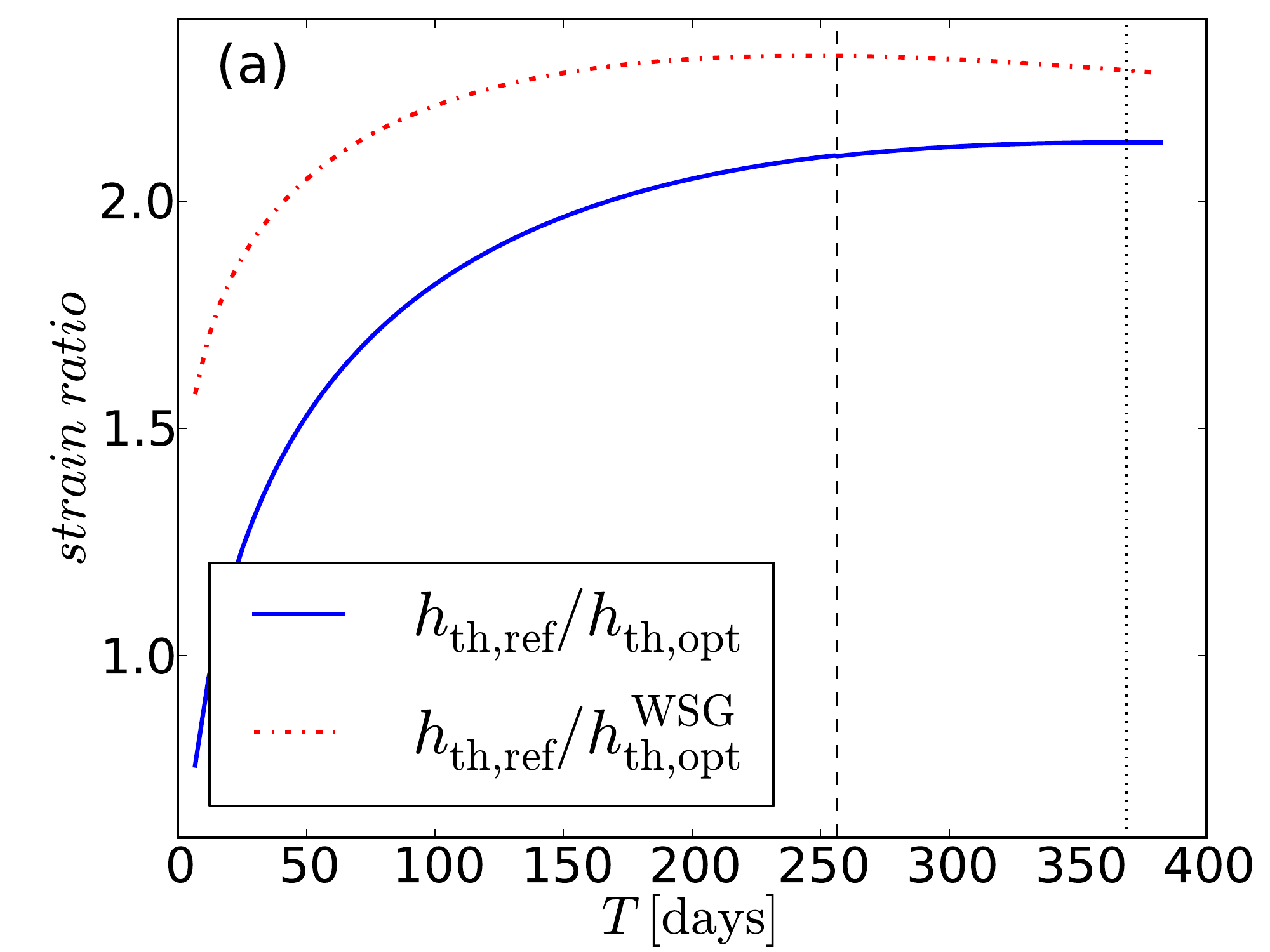}
  \includegraphics[width=0.8\columnwidth]{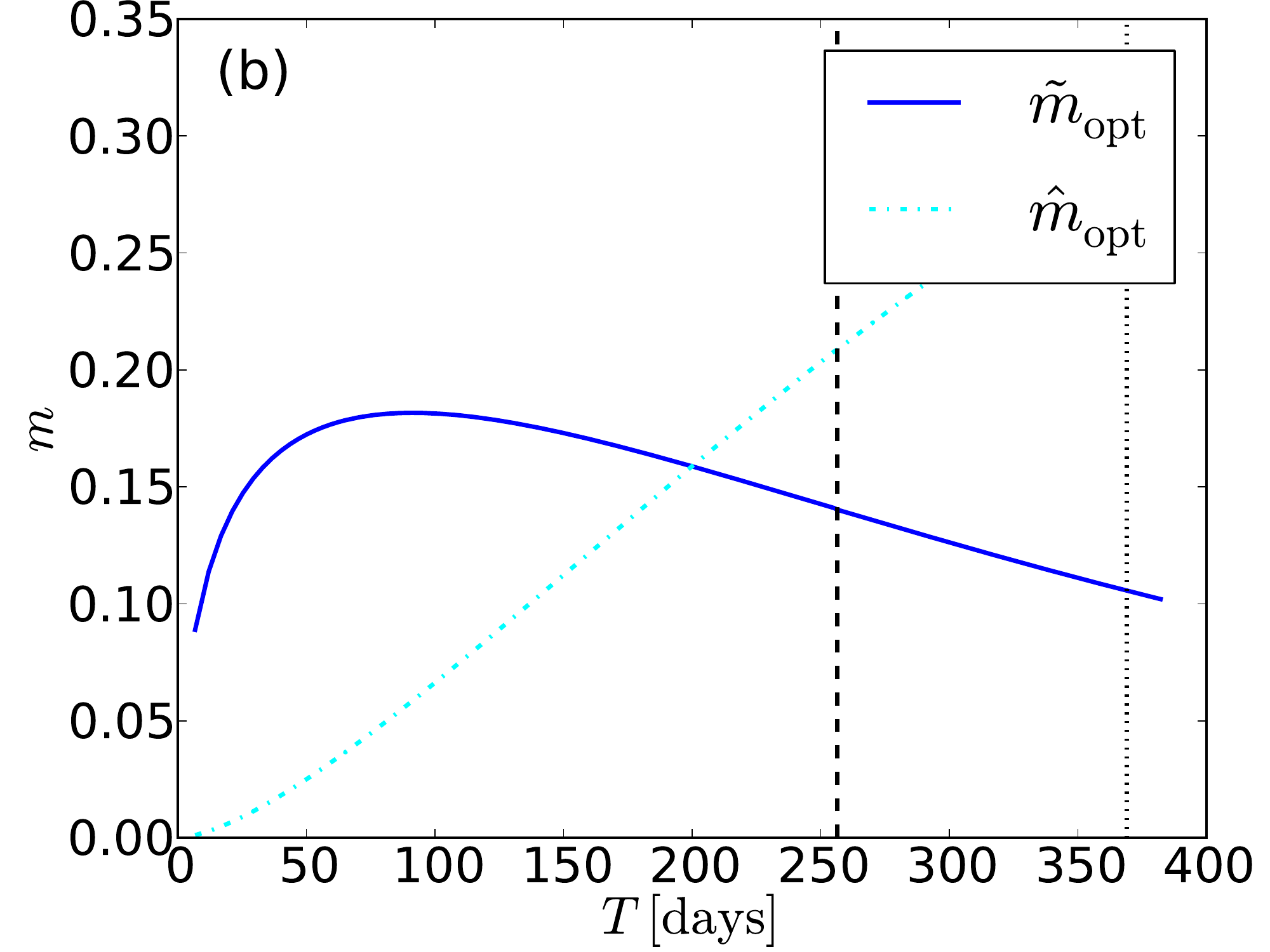}
}
\mbox{
  \includegraphics[width=0.8\columnwidth]{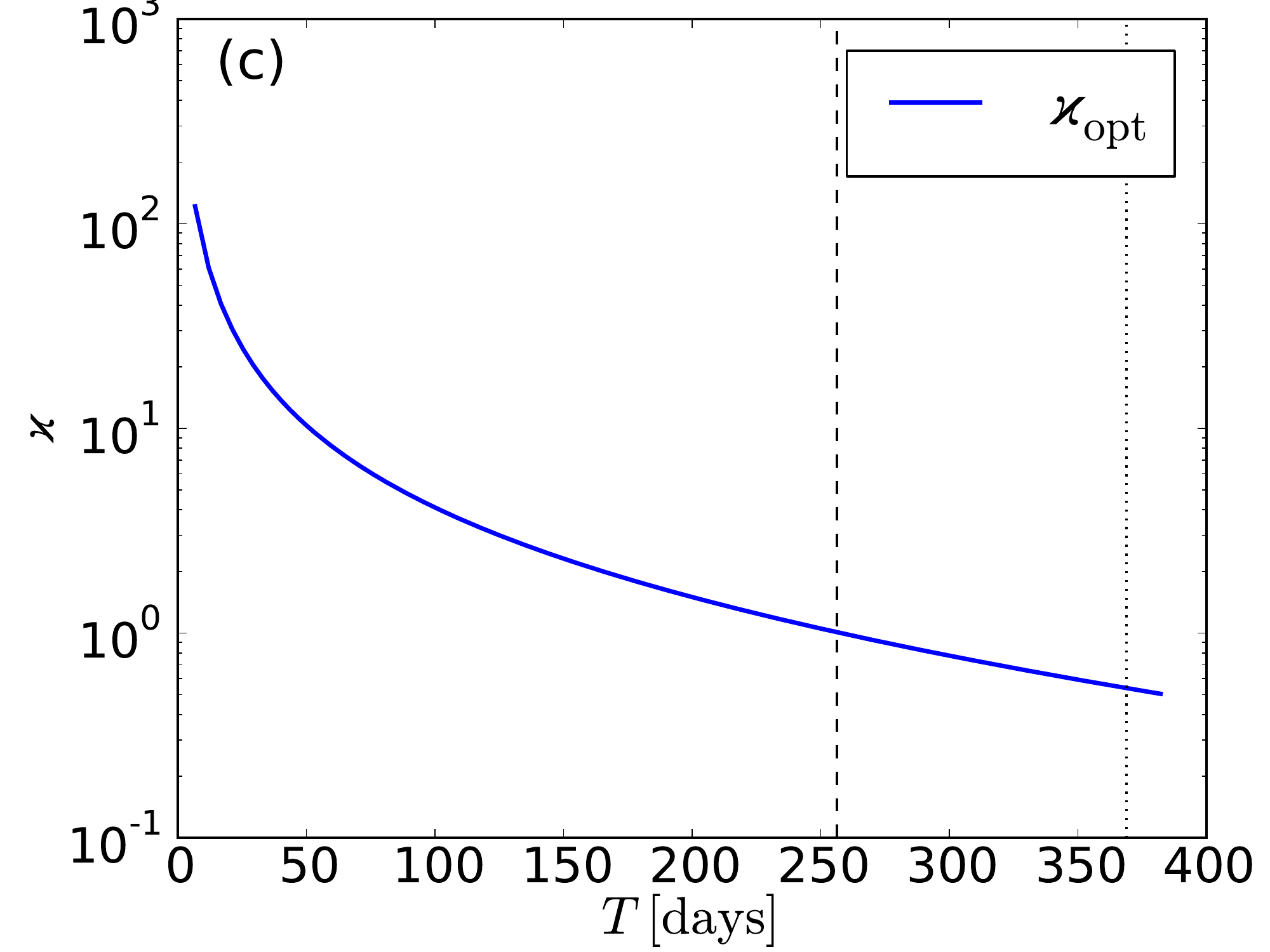}
  \includegraphics[width=0.8\columnwidth]{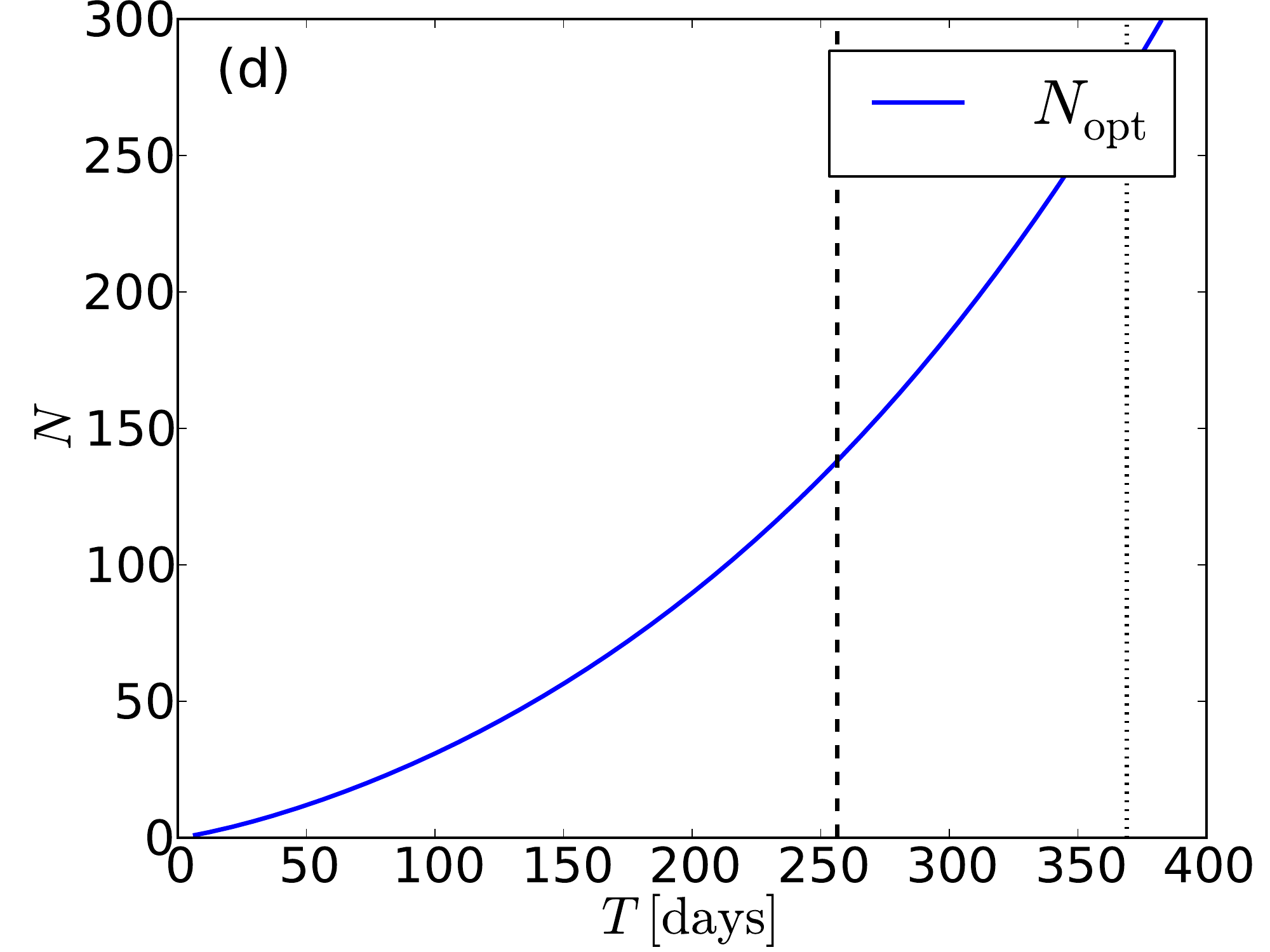}
}
\caption{Numerical optimal fixed-$\Tobs$ solution for a directed CasA search as a function of $\Tobs$.
  The dashed vertical line indicates the analytical WSG-optimal solution of Eq.~(\ref{eq:105}),
  while the dotted vertical line corresponds to the exact optimal solution.
  Panel (a) shows the weakest detectable signal strength compared to the reference value
  $h_{\thresh,\mathrm{ref}}$ of Eq.~(\ref{eq:97}), for the exact $\hth$ and for the WSG-approximated $\hth^\WSG$ (using $\dev=1$).
  (b) shows the optimal mismatch parameters $\opt{\miscmax}(\Tobs)$ and $\opt{\misfmax}(\Tobs)$,
  (c) shows the optimal computing-cost ratio $\opt{\cratio}(\Tobs)$ and
  (d) the optimal number of segments $\opt{\Nseg}(\Tobs)$ (treated as continuous).
}
\label{fig_casasentcoh}
\end{figure*}

Next we consider a StackSlide search over the same parameter space using the same computing cost $\CC_0$.
Assuming the optimal solution will have segment lengths in the range
$1\;\days \lesssim \Tseg \lesssim 7\; \days$, and a total span of $\Tobs \lesssim 365\,\days$, the
parameter-space dimensions would be $\nc = 2$, $\nf= 3$ (see \cite{2008CQGra..25w5011W}). This
results in power-law exponents $\deltac = 4$, $\deltaf = 6\,,\;\etaf = 4$, and therefore $\epsc=3$,
$\epsf=2$, and $\critc=2+6(\dev-1)$, $\critf = -2 + 4(\dev-1)$.
In order to simplify the example we use the WSG approximation, i.e.\ $\dev=1$, which implies that
the search would be bounded ($\critf<0$). We can therefore use Eq.~(\ref{eq:76}) to obtain the
optimal computing-cost ratio as
\begin{equation}
  \label{eq:85}
  \opt{\cratio} = 1\,.
\end{equation}
Note that when $\dev>1.25$ we would have $\critf>0$ and therefore this search would become unbounded.
From Eq.~(\ref{eq:59}) we obtain $\optO{\miscmax} \approx 0.29$, $\optO{\misfmax} \approx 0.55$,
and using Eq.~(\ref{eq:48}) we find the respective optimal mismatches as
\begin{equation}
  \label{eq:105}
  \opt{\miscmax} \approx 0.16 \,,\quad
  \opt{\misfmax} \approx 0.24\,.
\end{equation}
Using Eq.~(\ref{eq:121}) we can extract the  computing-cost coefficients $\kappac \approx 3.14\times 10^{-17}$ and
$\kappaf \approx 3.12\times 10^{-34}$ (with time measured in seconds), and plugging this into
Eqs.~(\ref{eq:100a}), (\ref{eq:100b}) we find the optimal StackSlide parameters as
\begin{equation}
  \label{eq:106}
  \begin{split}
    \opt{\Nseg} = 139\,,&\quad
    \opt{\Tseg} \approx 1.9\,\days\,,\\
    \opt{\Tobs} &\approx 266.5\,\days\,,
  \end{split}
\end{equation}
which is self-consistent with the initially-assumed template-bank dimensions, as it falls into the
assumed ranges for $\Tseg$ and $\Tobs$.

We can estimate the resulting sensitivity by solving Eqs.~(\ref{eq:97a}),(\ref{eq:97b}), which
yields $\RHOsumF^* \approx 17.3$, and substituting into Eq.~(\ref{eq:99}) we find a weakest
detectable signal strength $\hth$ of
\begin{equation}
  \label{eq:102}
  \left. \frac{\hth}{\sqrt{\Sn}} \right|_{\mathrm{opt}} \approx 2.4\times 10^{-3}\,\sqrt{\Hz}\,,
\end{equation}
which is an improvement on the optimal coherent sensitivity by more than a factor of {two}.

Figure~\ref{fig_casasentcoh} illustrates the behaviour of the optimal solution as a function of
$\Tobs$ without using the WSG approximation. This is obtained by numerically solving
Eq.~(\ref{eq:100b}) for $\opt{\cratio}(\Tobs)$, which yields $\opt{\mismax}(\Tobs)$ via
Eq.~(\ref{eq:48}) and $\opt{\Nseg}(\Tobs)$ via Eq.~(\ref{eq:100a}). We see that the non-WSG
approximated optimal solution results in somewhat different StackSlide parameters than the WSG
solution of Eq.~(\ref{eq:106}), but it hardly gains any further sensitivity.

Increasing the total computing cost $\CC_0$ would increase the relative
advantage of the StackSlide method compared to a fully-coherent search: the coherent search would
gain sensitivity as $\hth^{-1} \propto \CC_0^{1/14}$ according to Eq.~(\ref{eq:53c}), while the StackSlide
search would gain sensitivity as $\hth^{-1} \propto \CC_0^{1/10}$ according to Eq.~(\ref{eq:49}) (in the WSG approximation),
so here the StackSlide search is more ``efficient'' at converting increases of computing-power into
gains of sensitivity.

\subsection{All-sky CW search using Einstein@Home}
\label{eath}

As an example for a wide parameter space all-sky search with massive computing power,
we consider two recent CW searches performed on the Einstein@Home computing platform
\cite{EinsteinAtHome,2009PhRvD..79b2001A,Abbott:2009nc}, namely the StackSlide searches labelled
'S5GC1' and 'S6Bucket', which employed an efficient grid mapping implementation described in
\cite{2009PhRvL.103r1102P}.

An Einstein@Home search divides the total workload into many small \emph{workunits}, each of which covers a
small fraction of the parameter space and requires only a couple of hours to complete on a
host machine.
These searches consisted of roughly $10^7$ workunits. The E@H platform delivers a computing
power of order $100$~Tflop/s, and these searches ran for about 6~months each, so we can estimate
their total respective computing cost is of order $\CCtot\sim10^{21}\,$flop (i.e. $\sim1$ Zeta flop).
Each E@H workunit is designed to require about the same computing cost, which allows us to
base the present analysis on just a single workunit.

\begin{table*}[htbp]
  \centering
\begin{tabular}[c]{c| c|c|c|c || c|c|c|c || c|c|c}
&  $\deltac$ & $\deltaf$ & $\etaf$ & $\dev$ & $\Nseg$ & $\Tseg$[d] & $\miscmax$ & $\misfmax$ & $\CC_0$[h] & $\cratio$ & $\frac{\hth}{\sqrt{\Sn}}[\rtHz]$ \\\hline
S5GC1 &  $8.7$ & $7.7$ & $2.0$ & $1.1$ & $205$ & $1.0$ & $0.50$ & $0.50$ & $0.91$ & $2.545$ & $2.69\times 10^{-3}$\\
$\Tmax=1\,$y &  $10.0$ & $9.0$ & $2.0$ & $1.1$ & $528$ & $0.7$ & $0.14$ & $0.17$ & $0.91$ & $0.869$ & $2.19\times 10^{-3}$\\
\hline
S6Bucket &  $4.6$ & $3.6$ & $2.0$ & $1.2$ & $90$ & $2.5$ & $0.50$ & $0.50$ & $2.54$ & $13.914$ & $2.20\times 10^{-3}$\\
$\Tmax=1\,$y &  $3.7$ & $2.7$ & $2.0$ & $1.2$ & $175$ & $2.1$ & $0.58$ & $0.32$ & $2.54$ & $1.815$ & $1.93\times 10^{-3}$\\
\end{tabular}

  \caption{Einstein@Home example setups 'S5GC1' and 'S6Bucket', with corresponding results from an
    iterative optimization at fixed computing power $\CC_0$, with assumed maximal observation
    time of $\Tmax=1\,\years$. The gains in weakest detectable signal strength $\hth$
    are $\sim23\%$ and $\sim14\,\%$, respectively.}
  \label{tab:eah_results}
\end{table*}

The detector data used in these searches contained non-stationarities and gaps, and the template
banks were constructed in somewhat semi-empirical ways that are hard to model analytically.
In order to simplify this analysis we assume gapless stationary Gaussian data, and we
use the analytic metric expressions from \cite{Pletsch:2010a} to estimate the number of templates.
This example is therefore ``inspired by'' recent E@H searches, but does not represent a detailed
description of their computing cost or sensitivity.

The two searches 'S5GC1' and 'S6Bucket' covered a fixed spindown-range corresponding
to a spindown age of $\tau =
f_0/\dot{f} = 600\,\years$ at a reference frequency of $f_0=50\,\Hz$.
Each workunit covers a frequency-band of $\Delta f = 0.05\,Hz$, the spindown range of
$\Delta\fdot=2.7\times10^{-9}\,\Hz/\secs$ and a (frequency-dependent) fraction $q$ of the sky.
We can incorporate the sky-fraction $q$ by using template counts $q\Nt$ in the computing-cost
expressions, where $\Nt$ are the all-sky expressions from \cite{Pletsch:2010a}.
For simplicity we fix the parameter-space dimension to $n=4$, namely \{sky, frequency, spindown\},
and we use [Eq.~(56),(50) in \cite{Pletsch:2010a}]\footnote{There are missing terms in
  both [Eqs.~(57) and (83) in \cite{Pletsch:2010a}], but one can use their Eqs.~(50) and (84)
  instead to compute $\det g$.} for the number of coarse-grid templates $\Ntc$
and the refinement factor $\refac(\Nseg)$ of [Eq.~(77) in \cite{Pletsch:2010a}] (assuming gapless data).

For a workunit at $50\,\Hz$, the reference StackSlide parameters are:
\begin{itemize}
\item S5GC1: $q=1/3,\, \Nseg = 205,\, \Tseg=25\,\hours$
\item S6Bucket: $q=1/51,\,\Nseg=90,\,\Tseg=60\,\hours$
\end{itemize}
For both searches the mismatch distributions of the coarse- and fine-grid template
banks are not well quantified, so we simply assume hyper-cubic template banks
($\pacfactor=1/3$) with $\miscmax=\misfmax=0.5$, i.e.\ an average total mismatch of $\avg{\mis} = 1/3$.
Plugging these parameters into the template-counting formulae of \cite{Pletsch:2010a}, together with
the timing constants of Eq.~(\ref{eq:19}) from the Cas-A example, we find a reference
per-workunit computing cost of $\CC_0\approx0.91\,\hours$ for S5GC1, and $\CC_0\approx2.5\,\hours$ for S6Bucket.\footnote{The actual
  E@H workunits take about $6\,\hours$ to complete on a machine with these timings, but these
  setups included bigger refinement factors $\refac$ due to gaps in the data, and used rather different
  template-bank designs.}
Table~\ref{tab:eah_results} shows the estimated sensitivity for these reference searches assuming
the same false-alarm and false-dismissal probabilities as in the previous section.

We can apply the analytical optimal solution from Sec.~\ref{sec:optim-sens-stacksl} with
the extracted power-law coefficients at the reference StackSlide parameters found in
Table~\ref{tab:eah_results}.
This initially places us into the \emph{unbounded} regime (i.e.\ $\critf>0$) for both 'S5GC1'
and 'S6Bucket'.
We therefore expect to improve sensitivity by increasing $\Tobs$ until we hit the assumed upper
bound of $\Tmax = 1\,\years$, so we solve Eq.~(\ref{eq:100b}) for $\opt{\cratio}(\Tmax)$,
substitute into Eq.~(\ref{eq:48}) for $\{\opt{\miscmax},\opt{\misfmax}\}$ and obtain
$\opt{\Nseg}$ from Eq.~(\ref{eq:100a}).

In order to find a \emph{self-consistent} solution, we need to iterate this procedure:
we extract new power-law coefficients at the new solution, then re-solve until the parameters
converge to better than $1\%$ accuracy. In the case of the 'S5GC1' search, the converged solution
falls into the unbounded regime. In the case of 'S6Bucket' the converged solution falls into
the bounded regime, but with $\optO{\Tobs}>\Tmax$. The optimal observation time is therefore
$\opt{\Tobs}=1\,\years$ in both cases, and the resulting converged solutions and power-law
coefficients are given in Table \ref{tab:eah_results}.
We see that (under the present idealized conditions) we could gain $\sim23\%$ in detectable signal
strength $\hth$ compared to the 'S5GC1' setup, and $\sim14\%$ compared to the 'S6Bucket' setup.

\subsection{All-sky search examples from CGK }
\label{sec:comp-prev-stud}

The all-sky search examples studied in CGK~\cite{PhysRevD.72.042004} provide another interesting test case
for our optimization framework. CGK considered a multi-stage optimization, but
we can treat their first-stage result as a single-stage optimization problem at fixed given
computing cost.
CGK discussed four different cases, namely a search for either ``young'' (Y)
neutron stars ($\tau=f/\dot{f}=40\,\years$) or ``old'' (O) neutron stars ($\tau=10^6\,\years$), using
either a ``fresh-data'' (f) or ``data-recycling'' (r) mode (a distinction that is irrelevant for our
present purpose).
The optimized CGK StackSlide parameters and computing-cost constraints are found in
[Tables I-VIII in \cite{PhysRevD.72.042004}], and are summarized in Table
\ref{tab:cgk_results}.
For the sensitivity estimates we use the same false-alarm and false-dismissal probabilities as in
Sec.~\ref{sec:search-cassiopeia}.

Note that we expect our results to improve on the sensitivity of the CGK solution, as they
incorporated an \emph{ad-hoc} constraint of $\mismax=\miscmax = \misfmax$, and the total average
mismatch in [Eq.(46) in CGK] incorrectly included only the contribution from one template grid
instead of both, as discussed in Sec.~\ref{sec:real-stackslide}.

The functional form of the template-bank equations (originally from
BC~\cite{Brady:1998nj}) in the CGK computing-cost model [Eq.(53) in CGK] is not consistent with the
generic form of Eq.~(\ref{eq:9}) with respect to the mismatch scaling.
We therefore resort to extracting (potentially fractional) ``mismatch dimensions''
$\{\nc,\nf\}$ using Eq.~(\ref{eq:38}), in order to fully reproduce their computing-cost function
with the power-law model of Eq.~(\ref{eq:18}). The scaling parameters $\{\delta,\eta\}$ are extracted via
Eq.~(\ref{eq:33}) and $\dev$ from Eq.~(\ref{eq:88}). The resulting values are given in Table
\ref{tab:cgk_results}, assuming the FFT/resampling method for the $\F$-statistic calculations.

\begin{table*}[htbp]
  \centering
\begin{tabular}[c]{c|c|c| c|c|c|c || c|c|c|c || c|c|c}
& $\nc$ & $\nf$ & $\deltac$ & $\deltaf$ & $\etaf$ & $\dev$ & $\Nseg$ & $\Tseg$[d] & $\miscmax$ & $\misfmax$ & $\CC_0$[Zf] & $\cratio$ & $\frac{\hth}{\sqrt{\Sn}}[\rtHz]$ \\\hline
Y/r & $3.0$ & $4.0$ & $3.1$ & $6.0$ & $4.2$ & $1.7$ & $10$ & $2.6$ & $0.78$ & $0.78$ & $0.94$ & $0.071$ & $7.31\times 10^{-3}$\\
$\Nseg=\Nseg_{\mathrm{CGK}}$ & $3.0$ & $4.0$ & $3.1$ & $6.0$ & $4.2$ & $1.7$ & $10$ & $2.1$ & $0.17$ & $0.48$ & $0.94$ & $0.482$ & $6.38\times 10^{-3}$\\
\hline
Y/f & $3.0$ & $4.0$ & $3.1$ & $6.0$ & $4.2$ & $1.7$ & $9$ & $2.7$ & $0.78$ & $0.78$ & $0.82$ & $0.086$ & $7.42\times 10^{-3}$\\
$\Nseg=\Nseg_{\mathrm{CGK}}$ & $3.0$ & $4.0$ & $3.1$ & $6.0$ & $4.2$ & $1.7$ & $9$ & $2.2$ & $0.18$ & $0.46$ & $0.82$ & $0.526$ & $6.50\times 10^{-3}$\\
\hline
O/r & $2.5$ & $2.5$ & $3.0$ & $10.0$ & $7.4$ & $1.8$ & $8$ & $14.8$ & $0.35$ & $0.35$ & $0.74$ & $0.004$ & $2.61\times 10^{-3}$\\
$\Nseg=\Nseg_{\mathrm{CGK}}$ & $2.8$ & $2.5$ & $3.0$ & $10.0$ & $7.4$ & $1.8$ & $8$ & $14.6$ & $0.03$ & $0.34$ & $0.74$ & $0.090$ & $2.46\times 10^{-3}$\\
\hline
O/f & $2.5$ & $2.6$ & $3.0$ & $10.0$ & $7.3$ & $1.7$ & $9$ & $11.8$ & $0.21$ & $0.21$ & $0.35$ & $0.009$ & $2.65\times 10^{-3}$\\
$\Nseg=\Nseg_{\mathrm{CGK}}$ & $2.7$ & $2.5$ & $3.0$ & $10.0$ & $7.3$ & $1.7$ & $9$ & $12.4$ & $0.04$ & $0.33$ & $0.35$ & $0.107$ & $2.56\times 10^{-3}$\\
\end{tabular}

  \caption{CGK example search setups for young ('Y') and old pulsars ('O'), using either fresh ('f')
    or recycling ('r') data-modes. The first line of each example gives the
    original CGK solution with respective extracted power-law coefficients, and the second line
    shows our optimal self-consistent solution with constraint $\Nseg=\Nseg_{\mathrm{CGK}}$.
    The computing cost $\CC_0$ is measured in Zeta-flop (1Zf $= 10^{21}$flop).
  }
  \label{tab:cgk_results}
\end{table*}

Using the extracted scaling coefficients to compute the optimal solution from
Sec.~\ref{sec:optim-sens-stacksl} results in a solution that is inconsistent with the initially
extracted scaling coefficients. An iteration over solutions, allowing both $\Nseg$ and $\Tobs$ to
vary, did not converge.
We therefore solve a simpler problem by fixing the number of segments to the original CGK
values, i.e.\ we constrain the solutions to $\Nseg=\Nseg_{\mathrm{CGK}}$.
We proceed by solving Eq.~(\ref{eq:100a}) for $\opt{\cratio}(\Nseg)$, closing the solution via
Eqs.~(\ref{eq:48}) and (\ref{eq:100b}). We then extract new power-law coefficients at this solution
and iterate this procedure until convergence to better than $1\,\%$ accuracy is achieved.
The resulting fixed-$\Nseg$ optimal solutions are given in Table \ref{tab:cgk_results}.
The respective improvements of the weakest detectable signal strength $\hth$ compared to the original
CGK solutions are $\sim15\,\%$ in the young (Y) pulsar case, and $~\sim5\,\%$ in the old (O) pulsar
case.

\subsection{CWs from binary neutron stars}

For CWs from NSs in binary systems with known sky-position (such as Sco-X1 and
other LMXBs), the search parameter space typically consists of the intrinsic signal frequency and
orbital parameters of the binary system, i.e.\ (projected) semi-major axis, orbital period $P$,
periapse angle, eccentricity and eccentric anomaly. The corresponding template-counting
formulae were initially studied in \cite{Dhurandhar:2000sd} for coherent searches.
These have recently been extended to semi-coherent searches by Messenger \cite{2011arXiv1109.0501M},
giving explicit template scalings in two limiting cases, namely
(i) \emph{short} coherent segments compared to the orbital period, i.e.\ $\Tseg \ll P$ , and
(ii) \emph{long} coherent segments, i.e.\ $\Tseg \gg P$.

\subsubsection*{(i) Short coherent segments ($\Tseg \ll P$)}
\label{sec:short-coher-segm}

One can change parameter-space coordinates and Taylor-expand in small $\Tseg/P\ll 1$
to obtain the coherent template scaling [Eq.~(24) in \cite{2011arXiv1109.0501M}]:
\begin{equation}
  \label{eq:51}
  \co{\Nt}_{\nc} \propto \Tseg^{\nc(\nc+1)/2}\,,
\end{equation}
where $\nc$ is the effective coherent parameter-space dimension using the new coordinates.
The coherent cost power-law coefficients are therefore $\deltac = \nc(\nc+1)/2 + \dl$ and $\etac=1$.

The semi-coherent template scaling including eccentricity results in a 6-dimensional
semi-coherent template bank, i.e.\ $\nf=6$, and a template scaling [Eq.~(28) in
\cite{2011arXiv1109.0501M}] of $\ic{\Nt}_{\mathrm{ecc}} \propto \Nseg\,\Tseg^7$.
In the case of small eccentricity one has $\nf=4$, and the template scaling given in [Eq.~(29)
in \cite{2011arXiv1109.0501M}] is $\ic{\Nt}_{\mathrm{circ}} \propto \Nseg\,\Tseg^5$.
In both cases the semi-coherent power-law exponents satisfy $\deltaf \ge 5$, and $\etaf=2$,
resulting in the critical parameter $\critf > 0$. This implies that the boundedness-condition
Eq.~(\ref{eq:101}) is always violated, i.e.\ one should use all the available data.

\subsubsection*{(ii) Long coherent segments ($\Tseg \gg P$)}
\label{sec:long-coher-segm}

In this limit the template scalings in both the coherent and semi-coherent step are
[Eqs.~(32,33) in \cite{2011arXiv1109.0501M}]: $\co{\Nt} \propto \ic{\Nt} \propto \Tseg^2$,
which is unusual as there is \emph{no refinement}. Therefore $\etac = \etaf = 1$,
and $\deltac=2+\dl$, while $\deltaf =
2$, and therefore $\epsf=1$. We see that always $\critc > 0$ and $\critf = 2(\dev-1)>0$, and
therefore binary-CW searches in the long-segment limit also fall into the \emph{unbounded} regime,
i.e.\ one should use all the data.

\section{Discussion}
\label{disc}

We have derived an improved estimate of the StackSlide sensitivity scaling, correctly accounting
for the mismatches from both coarse- and fine-grid template banks, which had been overlooked by
previous studies.
By locally fitting sensitivity and computing-cost functions to power laws we are able to
derive fully analytical self-consistency relations for the optimal sensitivity at fixed computing cost.
This solution separates two different regimes depending on the critical parameter $\critf$ of
Eq.~(\ref{eq:101}): a bounded regime with a finite optimal $\optO{\Tobs}$, and an unbounded regime
where $\optO{\Tobs}\rightarrow\infty$.

Several practical examples are discussed in order to illustrate the application of this framework.
The corresponding sensitivity gains in terms of the weakest detectable signal strength $\hth$
are found to be $\sim100\%$ compared to a fully coherent directed search for CasA, and about
$5\%-20\%$ compared to previous StackSlide searches such as Einstein@Home and the examples given in
CGK \cite{PhysRevD.72.042004}.
We show that CW searches for binary neutron stars seem to generally fall into the unbounded
regime where all the available data should be used irrespective of available computing power.

This study only considered single-stage StackSlide searches on Gaussian stationary gapless
data from detectors with identical noise-floors. Further work is required to extend this analysis to
more realistic data conditions.

\section{Acknowledgments}
This work has benefited from numerous discussions and comments from colleagues, in particular Holger
Pletsch, Karl Wette, Chris Messenger, Curt Cutler, Badri Krishnan and Bruce Allen.
MS gratefully acknowledges the support of Bruce Allen and the IMPRS on Gravitational Wave Astronomy
of the Max-Planck-Society.
This paper has been assigned LIGO document number \dcc.

\bibliography{gwdaw_2010_pro}

\end{document}